\documentclass[preprint2,tighten]{aastex6}
\usepackage{amsmath,amstext,amssymb}
\usepackage{footnote}
\usepackage[T1]{fontenc}
\usepackage{apjfonts}
\usepackage[all]{hypcap} %Links go to figures. This breaks deluxetables; use \capstartfalse \capstarttrue around deluxetables to fix it.
\usepackage[normalem]{ulem}

\newcommand{\hmsun}{{h^{-1}}{\rm M}_{\solar}}
\newcommand{\solar}{_{\mathord\odot}}
\newcommand{\hmpc}{\ifmmode{h^{-1}{\rm Mpc}}\;\else${h^{-1}}${\rm Mpc}\fi}
\newcommand{\hpc}{\ifmmode{h^{-1}{\rm pc}}\;\else${h^{-1}}${\rm pc}\fi}
\newcommand{\hMpc}{\ifmmode{h^{-1}{\rm Mpc}}\;\else${h^{-1}}${\rm Mpc}\fi}
\newcommand{\hGpc}{\ifmmode{h^{-1}{\rm Gpc}}\;\else${h^{-1}}${\rm Gpc}\fi}
\newcommand{\hkpc}{\ifmmode{h^{-1}{\rm kpc}}\;\else${h^{-1}}${\rm kpc}\fi}

\newcommand{\msun}{{\rm M}_{\solar}}

\newcommand{\mr}{\ifmmode{M_r}\;\else$M_r$\fi}
\newcommand{\rd}{\ifmmode{R_\delta}\;\else$R_\delta$\fi}
\newcommand{\ngals}{\ifmmode{N_{\rm gals}}\;\else$N_{\rm gals}$\fi}

\newcommand{\mvir}{M_{\rm vir}}
\newcommand{\vvir}{V_{\rm vir}}
\newcommand{\rvir}{R_{\rm vir}}

\newcommand{\kmps}{\,\rm{km}\,\rm{s}^{-1}}

\shorttitle{}
\shortauthors{Lu et al.}

\begin{document}

\title{The importance of preventive feedback: inference from observations of the stellar masses and metallicities of Milky Way dwarf galaxies}

\author{Yu~Lu\altaffilmark{1}}
\author{Andrew~Benson\altaffilmark{1}}
\author{Andrew~Wetzel\altaffilmark{1,2,3}$\dagger$}\thanks{$\dagger$ Caltech-Carnegie Fellow}%\altaffiliation{Caltech-Carnegie Fellow}
\author{Yao-Yuan~Mao\altaffilmark{4}}
\author{Stephanie~Tonnesen\altaffilmark{1}}
\author{Annika~H.~G.~Peter\altaffilmark{5}}
\author{Michael~Boylan-Kolchin\altaffilmark{6}}
\author{Risa H. Wechsler\altaffilmark{7}}
\altaffiltext{1}{The Observatories, The Carnegie Institution for Science, 813 Santa Barbara Street, Pasadena, CA 91101, USA}
\altaffiltext{2}{TAPIR, California Institute of Technology, Pasadena, CA 91125, USA}
\altaffiltext{3}{Department of Physics, University of California, Davis, CA 95616, USA}
\altaffiltext{4}{Department of Physics and Astronomy and the Pittsburgh Particle Physics, Astrophysics and Cosmology Center (PITT PACC), University of Pittsburgh, Pittsburgh, PA 15260, USA}
\altaffiltext{5}{CCAPP and Department of Physics, The Ohio State University, 191 W.\ Woodruff Ave., Columbus, OH 43210, USA;\\ Department of Astronomy, The Ohio State University, 140 W.\ 18th Ave., Columbus, OH 43210, USA}
\altaffiltext{6}{Department of Astronomy, The University of Texas at Austin, 2515 Speedway, Stop C1400, Austin, TX 78712-1205, USA}
\altaffiltext{7}{Kavli Institute for Particle Astrophysics and Cosmology; Department of Physics, Stanford University, Stanford, CA 94305, USA;\\ SLAC National Accelerator Laboratory, Menlo Park, CA, 94025, USA}

\begin{abstract}
Dwarf galaxies are known to have remarkably low star formation efficiency due to strong feedback.  
Adopting the dwarf galaxies of the Milky Way as a laboratory, we explore a flexible semi-analytic galaxy formation model to understand how the feedback processes shape the satellite galaxies of the Milky Way. 
Using Markov-Chain Monte-Carlo, we exhaustively search a large parameter space of the model and rigorously show that the general wisdom of strong outflows as the primary feedback mechanism cannot simultaneously explain the stellar mass function and the mass--metallicity relation of the Milky Way satellites. 
An extended model that assumes that a fraction of baryons is prevented from collapsing into low-mass halos in the first place can be accurately constrained to simultaneously reproduce those observations.
The inference suggests that two different physical mechanisms are needed to explain the two different data sets. 
In particular, moderate outflows with weak halo mass dependence are needed to explain the mass--metallicity relation, and prevention of baryons falling into shallow gravitational potentials of low-mass halos (e.g. "pre-heating") is needed to explain the low stellar mass fraction for a given subhalo mass. 
\end{abstract}


%%%%%%%%%%%%
\section{Introduction}\label{sec:introduction}

Our own galaxy, the Milky Way (MW), provides an excellent laboratory for constraining galaxy formation physics. 
In particular, the dwarf galaxies in the Milky Way system are excellent places to test feedback models because feedback is expected to be most effective in their shallow gravitational potential wells \citep[e.g.,][]{Dekel1986a, Thoul1996a, Benson2002a, Okamoto2010a}. 
Observations of kinematic properties of MW satellite galaxies have demonstrated that these dwarf galaxies are dominated by dark matter with mass-to-light ratios as high as $\sim100-1000$ \citep[e.g.,][]{Mateo1998a, Munoz2006a, Simon2007a}, which suggests that these satellite galaxies have low baryon fractions.
The question is then whether the baryons were lost through strong outflows or instead were never able to condense into those galaxies. 
In this paper, we attempt to distinguish between these two scenarios using a flexible Semi-Analytic Model (SAM) built on merger trees extracted from high-resolution zoom-in simulations of MW-size halos \citep{Mao2015a}. 
Understanding how feedback works is important not only for understanding galaxy formation but also for constraining the properties of dark matter, as the kinematic effect of the outflow can influence the distribution of dark matter \citep[e.g.,][]{Pontzen2012a, Sawala2016a}, which interferes with observational tests of dark matter models using astronomical data \citep[e.g.,][]{Maccio2010b, Parry2012a, Chau2017a}.

Previous theoretical studies have suggested that galaxy formation in small halos can be strongly affected by outflows powered by the injection of supernova energy into the gas \citep{White1978a, Dekel1986a} and photo-ionization heating \citep{Couchman1986a, Efstathiou1992a, Thoul1996a, Mo1996c, Bullock2000a}.
Implementing these processes in greater detail in SAMs, many authors have shown that these physical processes are crucial for reproducing observational properties of dwarf galaxies such as the abundance of the classical dwarfs of the Milky Way system and properties of low-mass galaxies in the field \citep[e.g.,][]{Kauffmann1993b, Benson2002a, Somerville2002a, Maccio2010a, Guo2011c}. 
More recently, \citet{Henriques2013a, Henriques2015a} explored the parameter space of the {\sc L-Galaxies} SAM and found that when the outflow mass from low-mass halos is kept out of the halo for a long period of time before being reincorporated back into the halo, the model could fit the evolution of galaxy luminosity function remarkably well. 
On the other hand, \citet{White2015a} found that a redshift-dependent mass-loading factor law can also match the evolution of galaxy stellar mass. 
Moreover, \citet{Hirschmann2016a} tested a number of models with various implementations of feedback, suggesting that the physics of feedback is not well understood.

Broadly speaking, the feedback processes considered important for the formation of low-mass galaxies can be classified into two categories: (1) \textit{ejective feedback}, which expels baryons from the galaxy to the intergalactic medium (IGM); and (2) \textit{preventive feedback}, which inhibits baryons from accreting onto galaxies or even their dark matter halos in the first place. 
In most galaxy formation models, ejective feedback is generally captured by outflows, which remove baryonic mass from the galaxy and deposit it into the IGM \citep[see][and references therein]{Benson2010c, Somerville2015a}.
A common form of preventive feedback considered in SAMs and hydro simulations for low-mass halos is photo-ionization heating (photoheating for short), which not only reduces radiative cooling but also prevents low-mass halos from accreting their full complement of baryons. 
Such effects of the reionization of the Universe have also been implemented in galaxy formation models by following a ``filtering mass'' proposed by \citet[][]{Gnedin2000a}. 
Indeed, more recent studies have demonstrated that preventive feedback stronger than what reionization is expected to provide might be also needed to solve certain problems in galaxy formation \citep[e.g.,][]{Mo2002a, Mo2005a, Lu2007a, Kauffmann2013a, Lu2015a, Christensen2016a}.

Nevertheless, our understanding of these different forms of feedback is still poor. 
One can gain deeper understanding of these feedback mechanisms by testing models against observational data. 
Recently, \citet{Hirschmann2016a} implemented various feedback prescriptions into independent SAMs and compared the model predictions of stellar mass, star formation rate, gas fraction, and metallicities of field galaxies with observational data \citep[also see][]{White2015a}. 
While the authors claimed that none of the tested models appeared to be completely satisfactory in reproducing all the adopted observational data, they found that the models with strong ejective and preventive feedback were largely degenerate, pointing out the importance of breaking the degeneracy between the two feedback scenarios.  
Importantly, the authors have found a common feature of the relatively more successful models in their studies, i.e., more than half of the baryons associated with low-mass halos are kept unavailable for cooling or star formation at high redshift. 
Also, the authors have shown that matching the evolution of the galaxy metallicity-mass relation is generally challenging for galaxy formation models.

It has been shown that the metallicity--luminosity relation of dwarf galaxies provides stringent constraints for models of supernova feedback and reionization \citep{Li2010a, Font2011a, Starkenburg2013a, Gomez2014a, Cousin2016a, Hou2016a}.
Using an analytic model, \citet{Lu2015d} has demonstrated that galaxy metallicities tightly constrain the upper limit for the strength of outflow from a galaxy. 
If the outflow is too strong, it will blow out too much metal mass to match the observed metallicity--stellar mass relation. 
Interestingly, \citet{Font2011a} found that it was not possible for the GALFORM model \citep{Cole2000a} to simultaneously match the luminosity function and the metallicity--luminosity relation with the default supernova feedback prescription used in previous implementations of the model, in which the efficiency of feedback increases with decreasing halo circular velocity as a steep power law. 
The authors suggested that the feedback efficiency should saturate for halos with circular velocity $v_{\rm circ} \leq 65 \kmps$, and very strong reionization, a combination of cosmic and local reionization with 100 per cent ionizing photon escape fraction, was needed to simultaneously match the stellar mass function and the mass--metallicity relation of MW satellite galaxies. 
By varying a few parameters in a SAM, \citet{Hou2014a} found that the slope of the metallicity--stellar mass relation is sensitive to the strength of supernova feedback and reionization. 
Comparing their model results with the observed slope of the metallicity--stellar mass relation, the authors suggested that the Local Universe was reionized earlier than the cosmic average and a moderate supernova feedback was needed. 
Using a more sophisticated SAM, \citet{Hou2016a} found that, in addition to the early reionization of the local universe, the strength of ejective feedback must evolve with redshift to match the local constraints.

These previous studies provide very useful insight into understanding feedback in galaxy formation. 
However, these studies were carried out using galaxy formation models with hand-tuned parameters. 
Owing to the nonlinear nature of galaxy formation, it is often hard to fully capture the impact of multiple processes by varying one parameter at a time \citep[e.g.,][]{Henriques2009a, Bower2010a, Lu2011b, Gomez2012a, Benson2014a}.
A conclusive assessment for a model family can be obtained only if one widely explores the parameter space and studies the collective behavior of an ensemble of models that can match basic observational constraints.
In this paper, we employ a Markov-Chain Monte-Carlo (MCMC) machinery that allows us to explore the parameter space of a flexible galaxy formation model \citep{Lu2011b}. 
We investigate what form of feedback mechanisms are needed to explain two key observational constraints from the Milky Way satellite galaxies: the stellar mass function from the compilation of \citet{McConnachie2012a} and the stellar-phase metallicity--stellar mass relation of \citet{Kirby2011a}. 
Our goal for this study is to use those two relationships observed for the Milky Way satellite galaxies to break the degeneracy in the model and shed light on understanding the nature of feedback. 
In particular, we explore an ejective feedback model, in which the effect of feedback is to eject baryons out of halos following star formation, and an extended model, in which a fraction of the baryonic mass is prevented from collapsing into halos in the first place.

The paper is organized as follows.
We describe the SAM, the constraining data, and the method of inference in \S\ref{sec:method}. 
The results of the study are presented in \S\ref{sec:results}. 
In detail, we present the posterior distributions of both the ejective and extended models in \S\ref{sec:posterior}, 
show the stellar mass function and stellar-phase metallicity--stellar mass relation predicted by the constrained models and compare them with observations in \S\ref{sec:model_fit}, 
and discuss the strength of preventive feedback inferred from the data in \S\ref{sec:strength}. 
Finally, we summarize the conclusions and discuss the uncertainties of the study in \S\ref{sec:conclusions}.

%%%%%%%%%%%%
\section{Methodology}\label{sec:method}

\subsection{The Merger Trees}
%%% merger tree
In this study, we adopt a suite of dark matter-only high-resolution zoom-in simulations of MW-size halos, introduced in \citet{Mao2015a}, to model the stellar mass function and the stellar mass--metallicity relation of MW dwarf galaxies. 
The final halo virial masses of the simulated halos are in the range of $M_{\rm vir}=10^{12.1\pm0.03}\msun$, where the virial mass definition follows \citet{Bryan1998a}. 
This halo mass range is consistent with many observational constraints of the halo mass of the MW \citep{Cautun2014b, Eadie2015a, Xue2008a, Gonzalez2013a}.
The mass resolution of the simulations is $3.0 \times 10^5\,\hmsun$ per particle. 
The softening length in the highest-resolution region is $170 \, \hpc$ comoving. 
The cosmological parameters are $\Omega_{\rm M} = 0.286$, $\Omega_{\Lambda} = 0.714$, 
$h = 0.7$, $\sigma_8 = 0.82$, and $n_s = 0.96$, as described in in \cite{Mao2015a}. 

The mass resolution of merger trees extracted from the simulation is sufficient for the model to accurately characterize the properties of galaxies with stellar masses as low as $\sim 10^4\msun$ (corresponding to halo mass $\sim 10^9\msun$). 
For more details about this suite of zoom-in simulations, please refer to \citet{Mao2015a}.
We have tested the numerical convergence of the model by artificially reducing the mass resolution and the temporal resolution of the merger tree. 
We find that reducing the mass resolution by a factor of 10 and the temporal resolution by a factor of 5 produces negligible effects on the predicted stellar mass, metallicity, and star formation histories of the model galaxies. 

\citet{Lu2016a} find that the whole set of the simulated halos covers a fairly large parameter space in terms of the halo mass-assembly history, concentration, and subhalo mass function. 
Some host halos do not contain sufficiently high-mass subhalos to host the Magellanic Clouds (MCs) and are therefore  inconsistent with the observed stellar mass function, even though the model adopts extreme star formation and feedback parameters to boost the stellar mass to subhalo mass ratio. 
From the whole set of 46 zoom-in simulation halos, we select 14 of them whose subhalo population contains at least two subhalos with $V_{\rm max}\geq 55\kmps{}$, making them close analogues of the Milky Way halo \citep{Lu2016a}.
The exclusion of other host halos that do not contain high-mass subhalos from this study effectively avoids the situation where the model never fits the observed MW satellite stellar mass function because of the lack of rare high-mass subhalos for hosting the MCs. 

\subsection{The SAM}
In this study, we continue to use the model adopted in \citet{Lu2016a}, in which we employ flexible parameterizations for the baryonic processes of galaxy formation to encompass a wide range of efficiency for star formation and feedback.
The prescriptions for the baryonic processes implemented in the SAM are detailed in \citet{Lu2014b}, where we have shown that, aided with MCMC optimization, the model accurately matches the local galaxy stellar mass function and performs well in predicting the stellar mass functions out to $z\sim 6$. 
In this paper, we focus our investigation on the stellar feedback model, which is expected to be most crucial for shaping low-mass galaxies. 
Therefore, we only briefly describe the prescriptions that affect the properties of low-mass galaxies in the model. 

% reionization
\subsubsection{Reionization}\label{sec:reionization}
Heating due to UV photoionization not only offsets cooling of halo gas, but also prevents low-mass halos from accreting their full complement of baryons. 
The suppression of baryonic accretion into halos has been modeled using analytic and numerical models \citep{Gnedin2000a, Hoeft2006a, Okamoto2008a, Noh2014a}.
\citet{Gnedin2000a} showed that the fraction of baryons that can collapse into halos of a given mass in the presence of a photoionizing background can be described in terms of the ``filtering mass'', $M_{\rm F}$. 
Halos less massive than $M_{\rm F}$ accrete less baryonic mass than the universal average. 
\citet{Gnedin2000a} parametrized the collapsed baryon fraction as a function of redshift and halo mass with the expression
\begin{equation}\label{equ:reion}
f_{\rm b,reion}(z, M_{\rm vir})= \frac{f_{\rm b}}{[1+0.26 M_{\rm F}(z)/M_{\rm vir}]^3}~,
\end{equation}
where $f_{\rm b}$ is the universal baryon fraction and $M_{\rm vir}$ is the halo virial mass in dark matter-only simulations (or, equivalently the total mass of dark matter and baryons at the cosmic baryon fraction). 
The filtering mass is a function of redshift, and this function depends on the re-ionization history of the Universe.  
\citet{Kravtsov2004b} provide a fitting formula for the
filtering mass as a function of the redshift at which the first HII regions begin to overlap ($z_{\rm overlap}$) and the redshift at which most of the medium is re-ionized ($z_{\rm reion}$). 
We use the fitting functions (B2) and (B3) from appendix B of \citet{Kravtsov2004b} to compute the initial fraction of baryons that can collapse as a function of halo mass and redshift, $f_{\rm b, reion}$, with two parameters fixed at $z_{\rm overlap}=11$ and $z_{\rm reion}=10$.
The choice of the parameters is consistent with current cosmological constraints \citep[e.g.][]{Komatsu2009a, Planck-Collaboration2016a} and the results of our paper are insensitive to the precise values of these parameters.

\subsubsection{Star Formation}
The model assumes that cold gas is distributed in an exponential disk with a scale radius $r_{\rm gas}$, and only that gas mass with a surface density higher than a certain threshold, $\Sigma_{\rm  crit}$, can form stars \citep[e.g.][]{Kennicutt1998a, Kennicutt2007a,  Bigiel2008a}.  
This cold gas mass available for forming stars is then
\begin{equation}\label{equ:sf_msf_lu}
m_{\rm sf}=M_{\rm cold}     \left\{1-\left[1+\ln \left(\frac{\Sigma_{\rm cold,0}} {\Sigma_{\rm crit}}\right)\right] \frac{\Sigma_{\rm crit}}{\Sigma_{\rm cold,0}}\right\}~,
\end{equation}
where $M_{\rm cold}$ is the total cold gas mass of the galaxy, and $\Sigma_{\rm cold, 0}=M_{\rm cold}/(2\pi r_{\rm gas}^2)$ is the central surface density of gas disk.
This parameterization has two uncertain parameters, $r_{\rm gas}$ and $\Sigma_{\rm crit}$, which are clearly degenerate. 
We rewrite the term ${\Sigma_{\rm cold,0}}/ {\Sigma_{\rm crit}}$ as 
\begin{equation}\label{equ:sig}
\frac{{\Sigma_{\rm cold,0}}} {\Sigma_{\rm crit}} = \frac{M_{\rm cold}}{2\pi r_{\rm d,0}^2 \Sigma_{\rm SF}}\,, 
\end{equation}
where we define 
\begin{equation}
r_{\rm d,0}=\frac{0.035}{\sqrt{2}}\rvir\,,
\end{equation}
with $R_{\rm vir}$ representing the virial radius of the halo, and the parameter $\Sigma_{\rm SF}$ is defined as 
\begin{equation}
\Sigma_{\rm SF}=\left(\frac{r_{\rm gas}}{r_{\rm d,0}}\right)^2 \Sigma_{\rm crit}\,,
\end{equation}
in units of $\msun\, {\rm pc}^{-2}$.
When we use Eq.\ref{equ:sig} to replace the term in Eq.\ref{equ:sf_msf_lu}, the uncertainties of two parameters are absorbed by one parameter $\Sigma_{\rm SF}$. 

We assume that star formation rate is proportional to the mass of star forming gas and inversely
proportional to the dynamical timescale of the disk, $\tau_{\rm d}={r_{\rm d,0}/\vvir}$, yielding
\begin{equation}\label{equ:sf_lu}
SFR=\alpha_{\rm SF}\frac{m_{\rm sf}}{\tau_{\rm d}}~,
\end{equation}
where $\alpha_{\rm SF}$ governs star formation efficiency.
It is clear that $\alpha_{\rm SF}$ is degenerate with $\Sigma_{\rm SF}$ because both of them affect the efficiency of converting gas into stars. 
We choose to fix  $\alpha_{\rm SF}=0.01$, but allow $\Sigma_{\rm SF}$ to vary, so that we do not need to specify two other model parameters, $\Sigma_{\rm crit}$ and $r_{\rm gas}$, which are degenerate when the cold gas distribution is not constrained \citep{Lu2014a}.

% outflow
\subsubsection{Ejective Feedback}
The photoheating effect of reionization is expected to be effective only for halos with a circular velocity $\lesssim 40\kmps$ \citep{Gnedin2000a, Hoeft2006a, Okamoto2008a}.
Other feedback processes that are effective on more massive halos are needed to explain the stellar mass function of field galaxies and MW dwarfs. 
The most widely adopted stellar feedback model is in the form of outflows. 
Following star formation, the explosion of core collapse supernovae and stellar winds from high-mass stars provide enough energy to heat up the cold gas in the disc and drive strong outflows to expel baryonic matter from the galaxy. 
In models, the strength of the outflow is parameterized by a mass-loading factor, $\eta$, which characterizes the mass flux expelled out of the galaxy per unit rate of conversion of cold gas mass into stars,
${\dot M}_{\rm out} = \eta \phi$, where ${\dot M}_{\rm out}$ is the outflow rate and $\phi$ is the star formation rate of a galaxy. 
Regardless of the physics governing the outflow, the outflow mass-loading factor is generally parameterized as a power-law function of halo circular velocity $V_{\rm c}$ as  
\begin{equation}\label{equ:massloading}
\eta = \alpha_{\rm LD} \left( \frac{V_{\rm c}}{220 \kmps} \right)^{-\beta_{\rm LD}}\\,
\end{equation}
where $\alpha_{\rm LD}$ and $\beta_{\rm LD}$ are model parameters to be constrained. 
The $\alpha_{\rm LD}$ parameter governs the normalization for the mass-loading factor for halos with a circular velocity $V_{\rm c}=220 \kmps{}$. 
The parameter, $\beta_{\rm LD}$, controls how rapidly the mass-loading factor varies with the halo circular velocity.  
In the literature, when $\beta_{\rm LD} = 2$, the corresponding feedback model is referred as ``energy-driven wind'', while when $\beta_{\rm LD}=1$, it is referred as ``momentum-driven wind''. 

The gas mass in the outflow can either be trapped in the host halo or leave the halo depending on the depth of the gravitational potential well and the velocity of the outflow. 
We introduce two parameters, $V_{\rm out}$ and $\beta_{\rm out}$, to describe the behavior. 
If the circular velocity of the halo is lower than $V_{\rm out}$, the gravitational potential well of the halo is too weak to keep the outflow gas in the halo and most of the outflow mass is expelled from the halo; 
if the circular velocity of the halo is higher than $V_{\rm out}$, only a smaller fraction of the outflow gas can leave the halo. 
Quantitatively, the model is parameterized as
\begin{equation}
f_{\rm ej} = \left[ 1+ \left( \frac{ V_{\rm c}}{V_{\rm out}} \right)^{\beta_{\rm out}} \right]^{-1}~.
\end{equation}
In the parameterization, the fraction of the outflow mass that is expelled from the halo decreases with increasing halo circular velocity when $\beta_{\rm out}>0$. 
The rest of the outflow gas mass is added into the halo gas, which can cool onto the central galaxy. 

In the model, we deposit the expelled mass into a reservoir $M_{\rm ej}$ that is tracked for each halo.
The mass in this reservoir can increase due to outflow of the galaxies hosted by the halo and can decrease when the outflow mass is reincorporated back into the halo. 
A general parameterization for rate of this reincorporation can be written as 
\begin{equation} \label{equ:reincorporation}
{\dot M}_{\rm ej} = - \gamma_{\rm RI}\frac{M_{\rm ej}}{\tau_{\rm dyn}}\\,
\end{equation}
where $\tau_{\rm dyn}$ is the dynamical timescale of the halo at the virial radius, and $\gamma_{\rm RI}$ is a free parameter characterizing the efficiency of the reincorporation. 
This parameterization is different from the parameterization found in \citet{Henriques2013a}, where low-mass halos have longer time delays for the ejected mass to be reincorporated back into the halo than high-mass halos, similar to the results obtained by \citet{Oppenheimer2010a} using hydrodynamic simulations. However, using recent simulations, \citet{Christensen2016a} have shown that the re-accretion times depend only very weakly on halo mass.
\citet{Angles-Alcazar2017a} find that re-accretion of outflow mass in high-resolution hydrodynamic simulations occurs over a broad range of timescales. 
The efficiency parameter, $\gamma_{\rm RI}$, takes into account the deviation from the halo dynamical timescale and the uncertainties. 
We note that the ejected mass reservoir also tracks the metal mass that is ejected with the outflow. 
The instantaneous metallicity of the outflow is assumed to be equal to the metallicity of the ISM. 
The metal mass is assumed to be perfectly mixed in the ejected mass reservoir.
We assume perfect mixing as well for metals in other components, including the ISM and the halo gas.

\subsubsection{Preventive feedback}
In addition to this ejective outflow model, we also hypothesize a preventive feedback model. 
Differing from the ejective feedback model, the preventive feedback prevents some amount of baryonic mass from being accreted into halos. 
Such prevention can be due to any feedback process that prevents a halo from accreting its cosmic baryon fraction, for example strong heating from the local ionizing radiation field in additional to the global ionizing background or from the kinetic energy of strong feedback generated at earlier epochs of the Universe that propagates to a large Lagrangian volume.
Regardless of the physics behind the prevention, we can capture the effect of preventive feedback using a phenomenological model in which the fraction of baryons that would collapse into a halo is a function of halo mass and redshift. 
We adopt the following parameterization 
\begin{equation}\label{equ:fb}
f_{\rm b,pr} = \frac{\exp\left(\gamma_{\rm pr} z\right)}{1+\left(\frac{M_{\rm pr}}{M_{\rm vir}}\right)^{\beta_{\rm pr}}}\,,
\end{equation}
where $\beta_{\rm pr}$ and $\gamma_{\rm pr}$ are parameters characterizing the mass and redshift dependence, respectively, 
and $M_{\rm pr}$ is a parameter characterizing the mass scale below which the prevention becomes important. 
We treat $M_{\rm pr}$, $\beta_{\rm pr}$, and $\gamma_{\rm pr}$ as free parameters to be constrained by data. 
In this model, at a given redshift the halo baryon fraction decreases with decreasing halo mass as $f_{\rm b}\propto M_{\rm vir}^{\beta_{\rm pr}}$ when the halo mass is significantly lower than $M_{\rm pr}$. 
This captures the effect that low-mass halos form a shallower gravitational potential well, and thus baryon accretion onto low-mass halos can be more strongly affected by any heating processes. 
For a constant halo mass, the baryon fraction scales with redshift as $f_{\rm b}\propto \exp(\gamma_{\rm pr} z)$. 
In the case where $\gamma>0$, the accretion fraction increases with redshift for a constant mass. 
This captures the effect that the prevention is weaker at early times because the heating is less efficient or that halos at high redshift generally have sufficient gas supply and can accrete baryons at a higher rate. 
In the model, the total baryonic mass, summing up all the components in stars, cold gas, and ejected material, is required to be smaller than $f_{\rm b} f_{\rm b,reion} f_{\rm b,pr} M_{\rm vir}$, unless the halo has already converted all its baryonic matter into cold gas or stars. 
Also, by definition, $f_{\rm b, pr}$ cannot be larger than unity. 
When the numerical value for $f_{\rm b, pr}$ exceeds 1 for a halo, it simply means that no prevention is at work, so it is set to unity for the halo.

% metallicity
\subsubsection{Metallicity}
In the model, we follow not only the mass of the baryonic matter in the hot halo, in the disk as cold gas and stars, and in the component that is ejected from the halo, but also the metal mass in each phase. 
Metals are produced in stellar evolution. 
In the model, we adopt the so-called ``instantaneous recycling approximation'' to treat metal production. 
For a unit cold gas mass forming stars, we assume a fraction of the mass $R$ is instantaneously returned back to the ISM and a fraction of the mass $p$ is turned into metal mass and mixed into the ISM, $\delta M_{\rm Z} = p \delta M_*$. 
Assuming a Chabrier IMF \citep{Chabrier2003a}, we adopt the mass return fraction to be $R=0.43$ and the total metal yield to be $p=0.03$ in this paper. 
The produced metals are mixed into the cold gas in the disc. 
The metallicity of the cold gas is simply defined as the ratio of total metal mass in the cold gas and the total cold gas mass, as $Z_{\rm cold} = M_{\rm Z, cold} / M_{\rm cold}$. 
New stars have the average metallicity of the cold gas at the time they formed. 
As star formation proceeds, more metal mass in the galaxy is locked into the stellar mass. 
Similarly, the stellar metallicity in the model is defined as the ratio of total metal mass in stars and the total stellar mass, as $Z_{*} = M_{\rm Z, *} / M_{\rm *}$. 
When stellar feedback drives outflow, the metallicity of the outflow is assumed to be equal to the metallicity of the cold gas in the galaxy, which is an assumption widely adopted in SAMs.
When the ejected baryons are reincorporated into the halo, the metals are also brought back into the halo and mixed with the halo gas.
The model computes the radiative cooling rate of the hot halo gas with a given metallicity by interpolating the cooling tables of \citet{Sutherland1993a} for varying temperature and metallicity.  
We have adopted the value for the Solar metallicity as 
$Z_{\odot} = 0.0134$ \citep{Asplund2009a} when we compare the model predictions with observations.

To summarize, Table \ref{tab:model} provides a brief summary of all of the model parameters in the ejective model and the extended model. 
We also list the prior ranges over which we allow the parameters to vary.

\subsection{Constraining Datasets}
%%% the data
In this paper, we explore how the stellar mass function (SMF) and 
the stellar mass--metallicity relation (MZR) of MW satellite galaxies constrain galaxy formation physics. 
We use MCMC to explore the parameter space and to sample the posterior distribution of the model parameters under the data constraints of the MW satellite galaxies. 

%%% the likelihood functions
%% stellar mass function
For the stellar mass function, we adopt the likelihood function defined in \citet{Lu2016a} to perform Bayesian inferences. 
The likelihood function is described by the negative binomial distribution as proposed by \citet{Boylan-Kolchin2010a} for describing the distribution of the subhalo mass function predicted by $N$-body simulations:
\begin{equation}
L(D_{\rm SMF}|\theta) = \Pi_{i} P(N_i | r, p_i) = \Pi_i \frac{ \Gamma(N_i+r)} {\Gamma(r) \Gamma(N_i+1) } p_i^r (1-p_i)^{N_i}\,,
\end{equation}
where $N_i$ is the observed number of satellite galaxies for a given stellar mass bin $i$ per MW halo; the two parameters $r$ and $p_i$ are determined by the model as $r= { 1 }/{ s^2_{\rm I}}$, $p_i={1}/(1 + s^2_{\rm I} \mu_i)$; $\mu_i$ is the expected number for the $i^{\rm th}$ mass bin predicted by the model. $s_{\rm I}$ is the fractional scatter from the intrinsic scatter, $\sigma_{{\rm I},i}$, with respect to the Poisson scatter, $\mu_i$, defined as $s_{\rm I}\equiv \sigma_{{\rm I},i}/\mu_i$. 
We note that the value of $s_{\rm I}$ may vary as a function of mass bin and can be simulated for any given model if a large number of merger trees are utilized. 
In this paper, however, we assume it is a constant $s_{\rm I}=0.25$, as we have shown in \citet{Lu2016a} that this parameter does not have a strong impact on the likelihood of the data given a predicted stellar mass function. 
For the observational data, we adopt the stellar masses and memberships of MW satellite galaxies compiled in 
\citet{McConnachie2012a} and only use the properties of the most massive 11 satellite galaxies ($M_*= 2.9\times10^5\msun$) to constrain the model. 
\citet{Tollerud2008a} have shown that the incompleteness of the MW satellite galaxy count becomes important only for fainter dwarfs with $M_{\rm v}>-7$ or $L<10^{5}L_{\odot}$ unless there is a significant low-surface-brightness population of satellites like Crater 2 \citep{Torrealba2016a} with $L>10^{5}L_{\odot}$. 
We restrict this study to only the bright end of the stellar mass function to avoid uncertainties in incompleteness corrections.

%% metallicity
We use the observational results of \citet{Kirby2013a} as the data constraint on the stellar mass--metallicity relation of MW satellite galaxies. 
We run the model to predict the stellar-phase metallicity of all satellite galaxies in all the MW hosts. 
To fairly compare the model predictions with observations, we only use the satellites with stellar mass higher than $10^3\msun$. 
Our resolution tests indicate that our simulations produce converged results for galaxies with stellar mass $\gtrsim 10^4\msun$. 
Nevertheless, including galaxies one order of magnitude lower in stellar mass does not affect the overall trend of the stellar mass--metallicity relation predicted by the model over a large range of stellar masses. 
We then use a $2^{\rm nd}$ order polynomial function to fit the relation between the predicted stellar-phase metallicity and stellar mass for all the simulated samples at $z=0$ as 
\begin{equation}\label{equ:metfit}
\log Z(M_*) = \log Z_0 + c_1 \log M_* + c_2 \log M_*^2\\, 
\end{equation}
where $M_*$ is the stellar mass predicted for satellite galaxies, and $Z$ is the stellar-phase metallicity of a galaxy. 
The fit is dominated by model galaxies with $M_*>10^4\msun$. 
Tests show that including galaxies slightly below the conservatory resolution limit does not affect the results. 
The stellar-phase metallicity of a galaxy in the SAM is computed as the fraction of the metal mass that is locked in long-lived stars.  
The best-fit model is used to represent the mean relation between $\log \bar{Z}_*$ and $\log M_*$, where $\log \bar{Z}_*$ is the expected logarithmic metallicity given stellar mass $M_*$. 
When the mean relation is determined for each model, we compute the standard deviation of the predicted model galaxies about the mean relation, $\sigma_{\rm z}$. 
We assume that the standard deviation is a constant over the stellar mass range probed ($10^3\msun<M_*<10^8\msun$), which seems a good approximation for the data \citep{Kirby2013a}. 
\citet{Kirby2013a} showed that 
\begin{equation}
\log \left(\frac{Z_*}{Z_{\odot}}\right) = 0.3\pm0.02 \log \left(\frac{M_*}{10^6 \msun}\right) + (-1.69 \pm 0.04),
\end{equation}
where the best fit slope, $a_{\rm obs}=0.3$ and intercept $b_{\rm obs}=1.69$ and their $1\,\sigma$ errors $\sigma_a=0.02$ and $\sigma_b=0.04$. 
We then assume that the likelihood for a MW satellite galaxy with stellar mass $M_*$ to have stellar-phase metallicity $Z_*$ follows a Gaussian function with the mean and the standard deviation determined by the model. 
Thus, the likelihood function can be written as
\begin{equation}
p(Z_* | M_*) = \exp\left\lbrace- \frac{\left[Z_* - \bar{Z}_*(M_*)\right]^2} {2\sigma_{\rm z}^2}\right\rbrace.
\end{equation}
We assume each observed galaxy is independent from one another when computing the likelihood. Hence, the joint likelihood for a given set of observed galaxies is simply \begin{equation}
L(D_{\rm MZR}|\theta) = \prod_i P(Z_{*,i} | M_{*,i}).
\end{equation}
We note that the data constraint for the mass-metallicity relation extends to much lower masses than the stellar mass function we use because counts of MW dwarfs with $M_\star <10^5\msun$ likely suffer strongly from incompleteness. 
In our model for the likelihood function of the mass-metallicity relation, we have assumed that the low-mass satellites detected are representative of the full population at given mass. 
This may not be true if, for example, the satellites missed due to incompleteness have systematically different metallicities because the undetected ones are distributed farther away from the MW and have different star formation histories (and thus metallicities) than the detected satellite population.
We have tested this by using several randomly selected models with different model parameters. 
We find that the predicted metallicity of satellite galaxies does not systematically change with the distance between the satellite and the MW halo center. 
The stellar mass is always the main factor driving the change in metallicity. 
Therefore, we expect that our approach based on the likelihood function model proposed here should yield robust results.

Finally, we multiply the likelihood for the stellar mass function and the likelihood for the mass--metallicity relation together under the assumption that these two datasets are independent, to arrive at the total likelihood
\begin{equation}
L(D|\theta) = L(D_{\rm SMF}) \times L(D_{\rm MZR}).
\end{equation}

%%%% MCMC
We use MCMC to sample the posterior probability density distribution. 
In addition to the previous MCMC result adopted in \citet{Lu2016a}, two runs for each of the two competing models are presented in this paper.  
A brief description of the model parameters and priors are listed in Table \ref{tab:model}. 
For a detailed explanation of these parameters, readers are referred to \citet{Lu2014b}. 
For each model, we run the MCMC for 20,000 iterations with 144 parallel chains using the differential evolution algorithm \citep{Ter-Braak2006a}. 
The convergence test is done with the Gelman--Rubin test \citep{Gelman1992a}, requiring the potential scale reduction factor $\hat{R}<1.2$.
After removing outliers and pre-burn-in states, we obtain $\sim 650,000$ posterior samples from the MCMC for each run.

%%%%%%%%%%%%%%%%%%%%%%%%%%%%%%%%%%%%%%%%%%%%%%%%%%%%%%%%%%%%%%%%%%%%%%%%%

\begin{table*}[htb!]
\centering
\begin{tabular}{c p{6cm} c c c}
%{p{2cm} p{8cm} p{2cm} p{2cm}}
\hline
\hline
notation & meaning of parameter & prior &  ejective model posterior & extended model posterior
\\ 
\hline
$\Sigma_{\rm SF}$ & 
combined parameter for gas disk size and the gas surface density threshold for star formation
%\footnotemark[1]  
& 
$\log\Sigma_{\rm SF}\in[-1, 3]$& 
$[1.74, 2.23]$ &
$[1.97, 2.88]$
\\

$\alpha_{\rm LD}$ &
normalization of the mass-loading factor (Eq.~\ref{equ:massloading})&
$\log\alpha_{\rm LD}\in [-12, 2]$&
$[-4.94, 0.475]$ &
$[-10.6, -2.35]$
\\

$\beta_{\rm LD}$ &
power-law index for the circular velocity dependence of the mass-loading factor (Eq.~\ref{equ:massloading})&
$\beta_{\rm LD} \in [-4, 15]$ &
$[2.33, 7.87]$ &
$[-2.33, 6.15]$
\\

$V_{\rm out}$ &
characteristic halo circular velocity [$\kmps$], below which all outflow mass leaves the host halo &
$\log V_{\rm out} \in [0.5, 2.5]$&
$[1.44, 2.15]$ &
$[0.808, 2.15]$
\\

$\beta_{\rm out}$ & 
steepness of the transition from total outflow for halos with $V_c\ll V_{\rm out}$ to no outflow for halos with $V_c \gg V_{\rm out}$ &
$\beta_{\rm out} \in [0, 8]$&
$[1.83, 6.83]$ &
$[1.30, 6.70]$
\\

$\gamma_{\rm RI}$ &
fraction of outflow mass reincorporated back into the halo (Eq.~\ref{equ:reincorporation})&
$\log \gamma_{\rm RI} \in [-3, 0]$&
$[-2.95, -2.56]$ &
$[-1.68, -0.353]$
\\

$M_{\rm pr}$ &
characteristic halo mass scale [$\msun$] in prevention feedback (Eq.~\ref{equ:fb})&
$\log M_{\rm pr} \in [8, 12]$&
-- &
$[10.4, 11.1]$
\\

$\beta_{\rm pr}$ &
parameter for the mass dependence in prevention feedback (Eq.~\ref{equ:fb})&
$\beta_{\rm pr} \in [0, 4]$&
-- &
$[1.76, 3.59]$
\\

$\gamma_{\rm pr}$ &
parameter for the redshift dependence in prevention feedback (Eq.~\ref{equ:fb})&
$\gamma_{\rm pr} \in [0, 4]$&
-- &
$[0.389, 1.76]$
\\
\hline % inserts single-line
comment
 &
 &
 &
poor fit to SMF &
good fit to SMF \& MZR \\
\hline % inserts single-line
\hline
\end{tabular}
\caption{Summary of the semi-analytic model parameters. The first column lists the notations of the free parameters in the SAM. The second column briefly explains the meaning of each free parameter. 
The third column lists the prior for each parameter in our inference. 
The fourth and the fifth column list the posterior bounds enclosing 67\% marginalized posterior probability for the ejective feedback model 
and the extended model, respectively. 
Both models are constrained to the stellar mass function and the mass--metallicity relation of MW dwarf galaxies.  The posterior distributions are shown in Figure \ref{fig:post_pr}.  \label{tab:model}}
%\sout{\footnotetext[1]{The model parameter is defined as $\Sigma_{\rm SF}=\left({2 r_{\rm gas}}/{0.035 R_{\rm vir}}\right)^2 \Sigma_{\rm SF,0}$ in units of $\msun\, {\rm pc}^{-2}$, where $r_{\rm gas}$ is the exponential radius of the cold gas disk and $\Sigma_{\rm SF,0}$ is the critical gas surface density, above which star formation can happen \cite[e.g.][]{Kennicutt1998a}. The model parameter absorbs the uncertainties in both the size of the gaseous disk and the threshold surface density of the cold gas for star formation.} } 
\end{table*}

%%%%%%%%%%%%%%%%%%%%%%%%%%%%%%%%%%%%%%%%%

\section{Results}\label{sec:results}

\subsection{Posterior distribution}\label{sec:posterior}

\begin{figure*}[htbp]
\begin{center}
\includegraphics[width=0.85\textwidth]{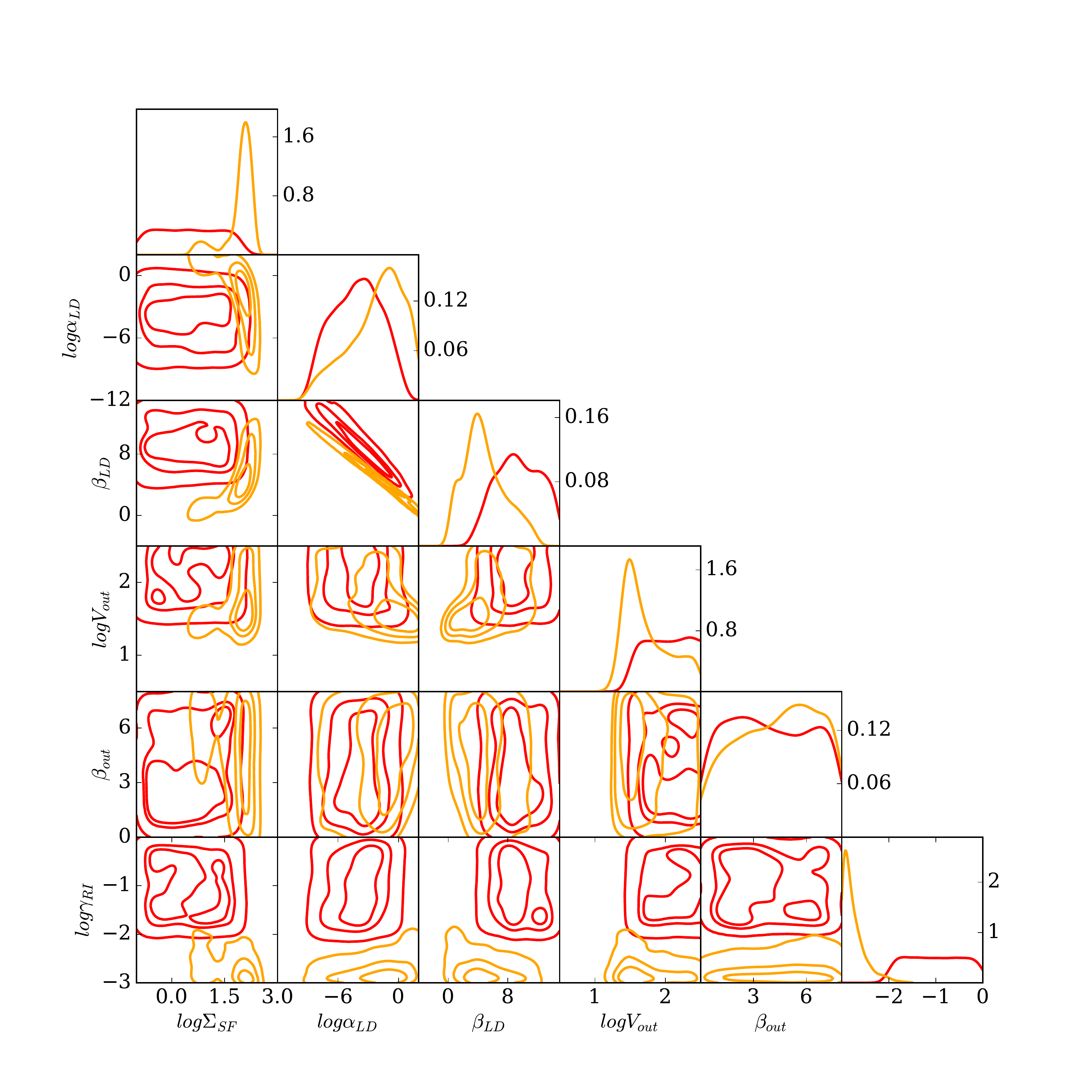}
\caption{The posterior distributions of the ejective feedback model constrained to different data. The red lines and contours show the 1-dimensional and 2-dimensional marginalized posterior probability distributions of the ejective model constrained to only the MW satellite stellar mass function. 
The yellow lines and contours show the same distributions of the same model constrained to both the MW satellite stellar mass function and the stellar-phase metallicity--stellar mass relation of MW satellite galaxies. 
}
\label{fig:post_ej}
\end{center}
\end{figure*}

\begin{figure*}[htbp]
\begin{center}
\includegraphics[width=0.85\textwidth]{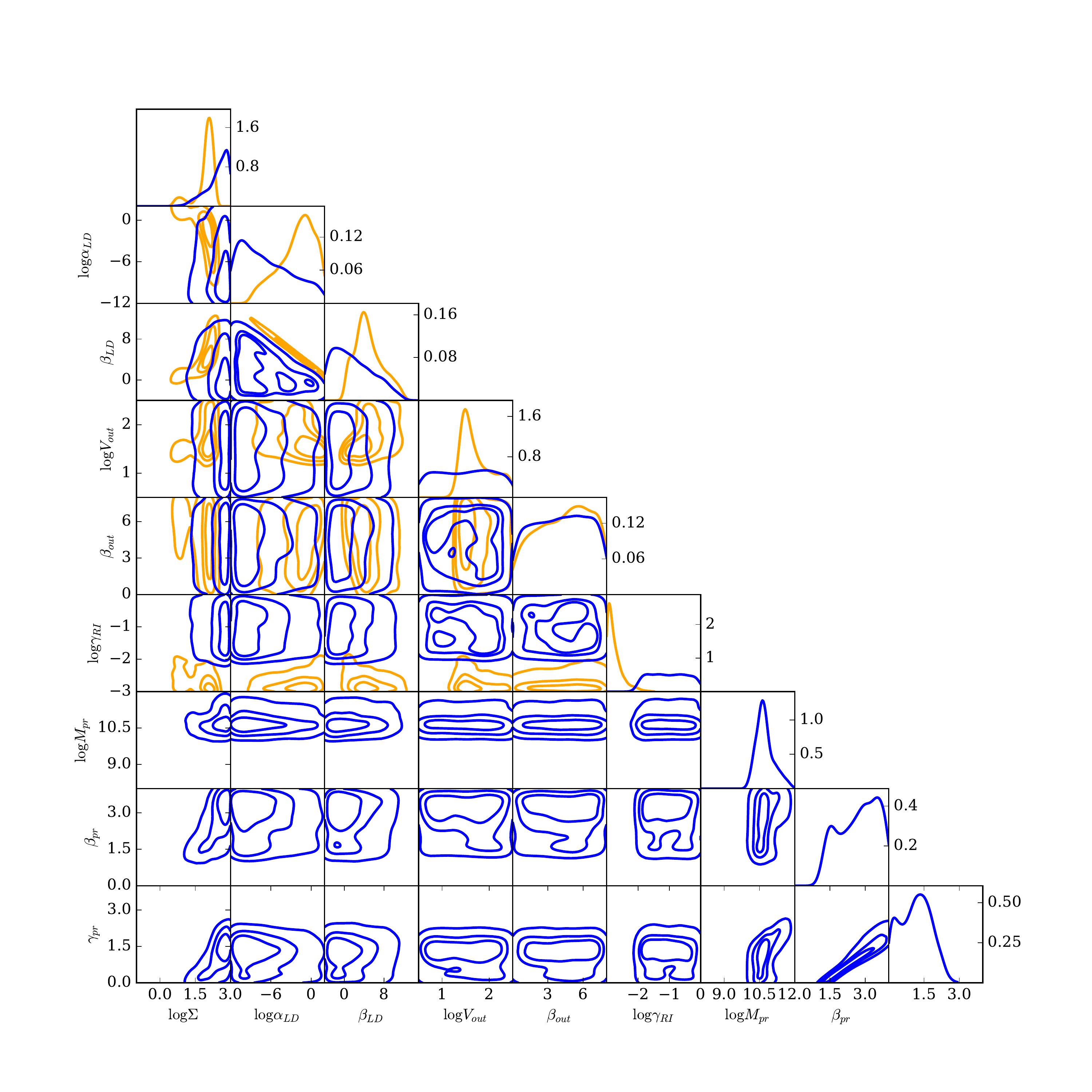}
\caption{The posterior distributions of two different models that are 
constrained to both the MW satellite stellar mass function 
and the stellar-phase metallicity--stellar mass relation. 
The blue lines and contours denote results of the extended model including both ejective and preventive feedback. 
The yellow lines and contours are the same as those shown in Fig. \ref{fig:post_ej}, denoting the ejective model constrained to the same data sets. 
}
\label{fig:post_pr}
\end{center}
\end{figure*}

We show the posterior probability distributions of the ejective model when it is constrained to different data in Figure \ref{fig:post_ej}. 
The panels on the diagonal line in the figure show the posterior distributions marginalized to each dimension of the parameters, and the off-diagonal panels show the 2-dimensional marginalized posterior probabilities. 
In each panel, the red lines or contours denote the posterior obtained using only the MW SMF as data constraint.
The same results have been presented in \citet{Lu2016a}, but we repeat the plot here to compare with new results.
The yellow lines and contours denote the new results constrained by both the MW SMF and the MZR. 
As one can see, the addition of the metallicity relation in the data strongly constrains the model. 
In the previous result with the stellar mass function only, the posterior distribution for many parameters were very broad. 
When the mass--metallicity relation is added into the constraints, the posterior becomes much narrower for almost all the free parameters and, more interestingly, some of the modes of the posterior distribution move around in the parameter space. 
We note that some of the posterior distributions are approaching the prior boundaries, potentially indicating that the prior should be further extended. 
We have tested varying the priors and found that further extending the priors does not change our conclusions.

We find that the addition of the mass--metallicity relation requires a different outflow mass-loading factor than when matching the stellar mass function alone. 
The normalization of the outflow mass-loading factor is constrained to higher values, 
but its halo circular velocity dependence is constrained to be weaker.  
This is reflected in the location of the modes in the contours of the $\log \alpha_{\rm LD}$ v.s. $\beta_{\rm LD}$ panel. 
From \citet{Lu2016a}, the stellar mass function constrained contours have a mode at $\beta_{\rm LD}\sim7$, but the mass--metallicity relation constrained contours have a mode at $\beta_{\rm LD}\sim 2$. 
As \citet{Lu2016a} have shown, the rapid increase in the mass-loading factor of the model results in significantly steeper MZR, underpredicting the stellar-phase metallicity for MW dwarfs. 
Although the normalization of the mass-loading factor in the newly constrained model is constrained to higher values, the mass-loading factor increases significantly more slowly as the halo circular velocity decreases, resulting in a much better fit to the MZR of the MW satellite galaxies. 
As we will discuss in the next subsection, however, the model fails to simultaneously fit both the mass--metallicity relation and the stellar mass function even when the parameters are allowed to vary in a broad range of prior. 

In addition, we find that the posterior distributions of other parameters are also altered to different degrees. 
The parameter $\Sigma_{\rm SF}$ is constrained to slightly higher values, indicating a higher cold-gas surface density or larger gas-disk size is preferred by the addition of the MZR in the data constraints. 
The parameter $V_{\rm out}$ is constrained to slightly lower values. 
Because halos with circular velocity higher than $V_{\rm out}$ retain outflow within the halo, this change indicates that the MZR prefers models that can retain outflow so the metal enriched gas can fall back on the galaxy.
For the gas mass ejected out of the halo, the newly constrained model prefers significantly lower $\gamma_{\rm RI}$, which means that the model is pushed to have very inefficient reincorporation to better match the data. 
In the next subsection, we will marginalize all the variations of the parameters in the analysis and assess the model in comparison with the data. 

Figure \ref{fig:post_pr} shows the posterior distribution of the extended model, which includes both ejective feedback and preventive feedback, constrained to both the MW satellite stellar mass function and the mass--metallicity relation. 
In comparison, we also include the posterior distribution of the ejective model constrained to the same data sets in the panels where the parameters are in common for both models. 
Since the last three parameters, $\log M_{\rm pr}$, $\beta_{\rm pr}$ and $\gamma_{\rm pr}$ are newly introduced in the extended model, there are no counterpart contours and distribution functions for those parameters from the ejective model. 
The marginalized posterior distribution for the three additional parameters have the following characteristics. 
First, the characteristic prevention mass scale is strongly constrained to $M_{\rm pr}\sim 10^{10.7}\msun$, indicating that halos with mass lower than this would accrete baryons at a lower rate than the halo mass accretion rate multiplied by the cosmic baryon fraction at late times.
Second, at a fixed redshift, the baryon accretion fraction governed by the preventive feedback scales with the halo mass as $\mvir^{\beta_{\rm pr}}$, where the parameter $\beta_{\rm pr}$ is constrained to a fairly broad distribution peaking at $\sim 3$.
Third, the characteristic prevention mass scale increases with increasing redshift as $\exp(\gamma_{\rm pr} z)$, where the parameter $\gamma_{\rm pr}$ is also constrained to a fairly broad distribution peaking at $\sim 1.4$ in strong degeneracy with varying $\beta_{\rm pr}$.
Recall that reionization is included in all of our models as discussed in Section \ref{sec:reionization}.
The fact that the parameters for the preventive feedback model are strongly constrained by the data suggests that preventive feedback in addition to reionization is needed to better fit the data. 

In addition to the features for the new parameters in the posterior distribution, we also find that when the preventive feedback model is included, the parameters for the outflow model are constrained to different values. 
First of all, the extended model requires weaker outflow to fit the data. 
This can be found by looking at the marginalized posterior probability distribution of parameter $\log \alpha_{\rm LD}$, which is the normalization of the mass-loading factor. 
This parameter is required to have high values for the pure ejective model, but is now constrained to have a much lower values with a more extended distribution within the prior range, indicating that only a moderate level of outflow is needed to explain the stellar mass function and the mass--metallicity relation. 
The scaling relation for the outflow mass-loading factor is also constrained to be a weaker function of halo circular velocity. 
This can be seen in the posterior probability distribution of parameter $\beta_{\rm LD}$.
Comparing the posterior distribution of this parameter for the two models, one can find that the peak of the distribution for $\beta_{\rm LD}$ in the extended model moves to lower values, $\beta_{\rm LD}\sim 0$ within a fairly broad range. 
We find that the ``energy-driven wind'' model ($\beta_{\rm LD}=2$) and the ``momentum-driven wind'' model ($\beta_{\rm LD}=1$) can explain the mass--metallicity relation fairly well when other parameters are tuned accordingly.  
The results suggest that when a certain fraction of baryonic matter is prevented from collapsing into halos in the first place, a large range of possible outflow models with lower mass-loading factors can reproduce the stellar mass function and the mass--metallicity relation. 
Moreover, other parameters are also changed accordingly. 
The parameter $\Sigma_{\rm SF}$ characterizing the cold-gas surface density threshold for star formation and the size of the cold-gas disk is constrained to higher values. 
The transition halo circular velocity, $V_{\rm out}$, now has a flat distribution extending to very low values, indicating that the model prefers that outflow gas is retained in the halo rather than leaving the halo. 
For the outflow gas that leaves the halo, the larger values of $\gamma_{\rm RI}$ indicate a more rapid reincorporation.

\subsection{Model fit}\label{sec:model_fit}

\begin{figure*}[htbp]
\begin{center}
\includegraphics[width=0.3\textwidth]{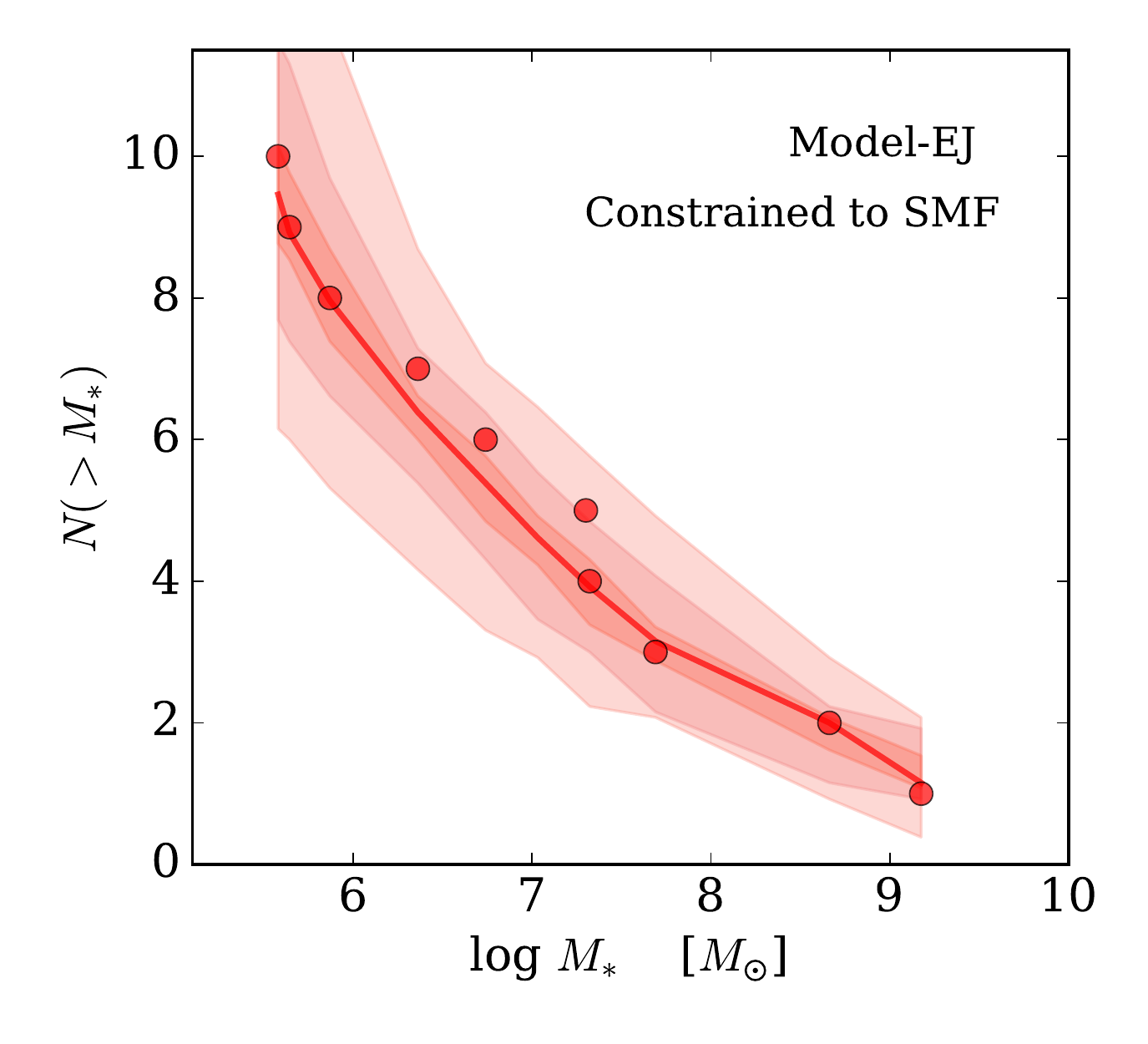}
\includegraphics[width=0.3\textwidth]{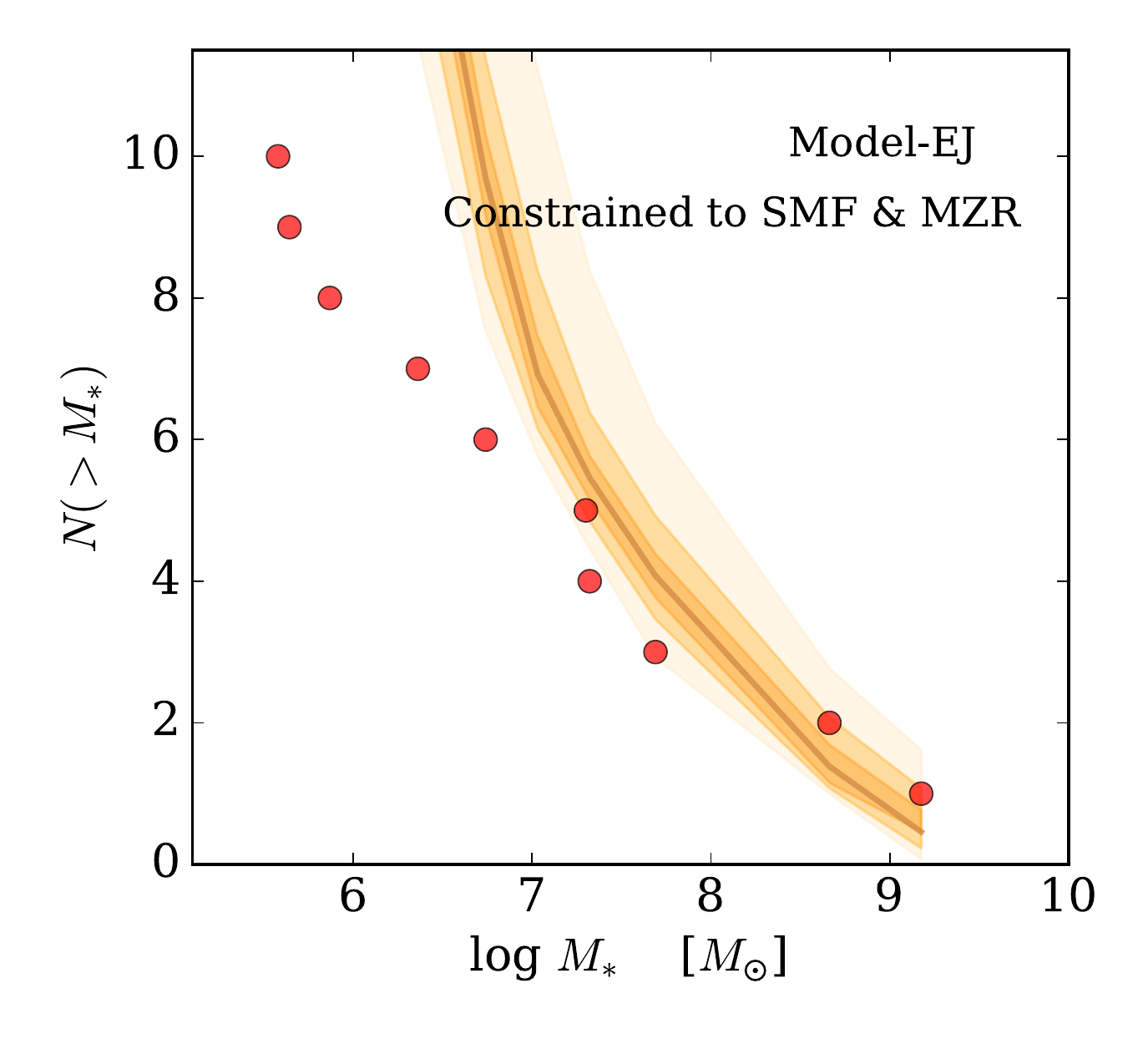}
\includegraphics[width=0.3\textwidth]{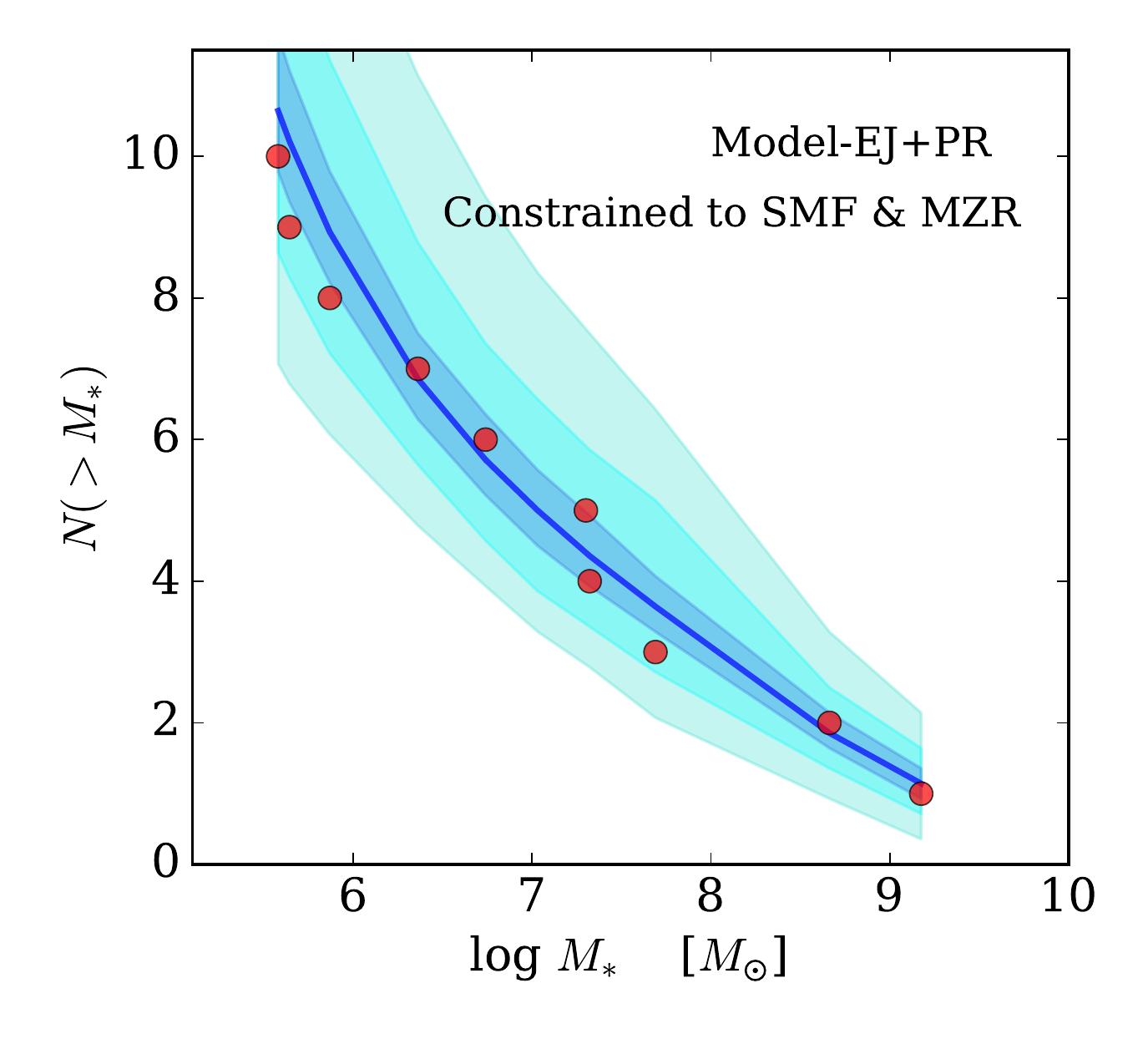}
\includegraphics[width=0.3\textwidth]{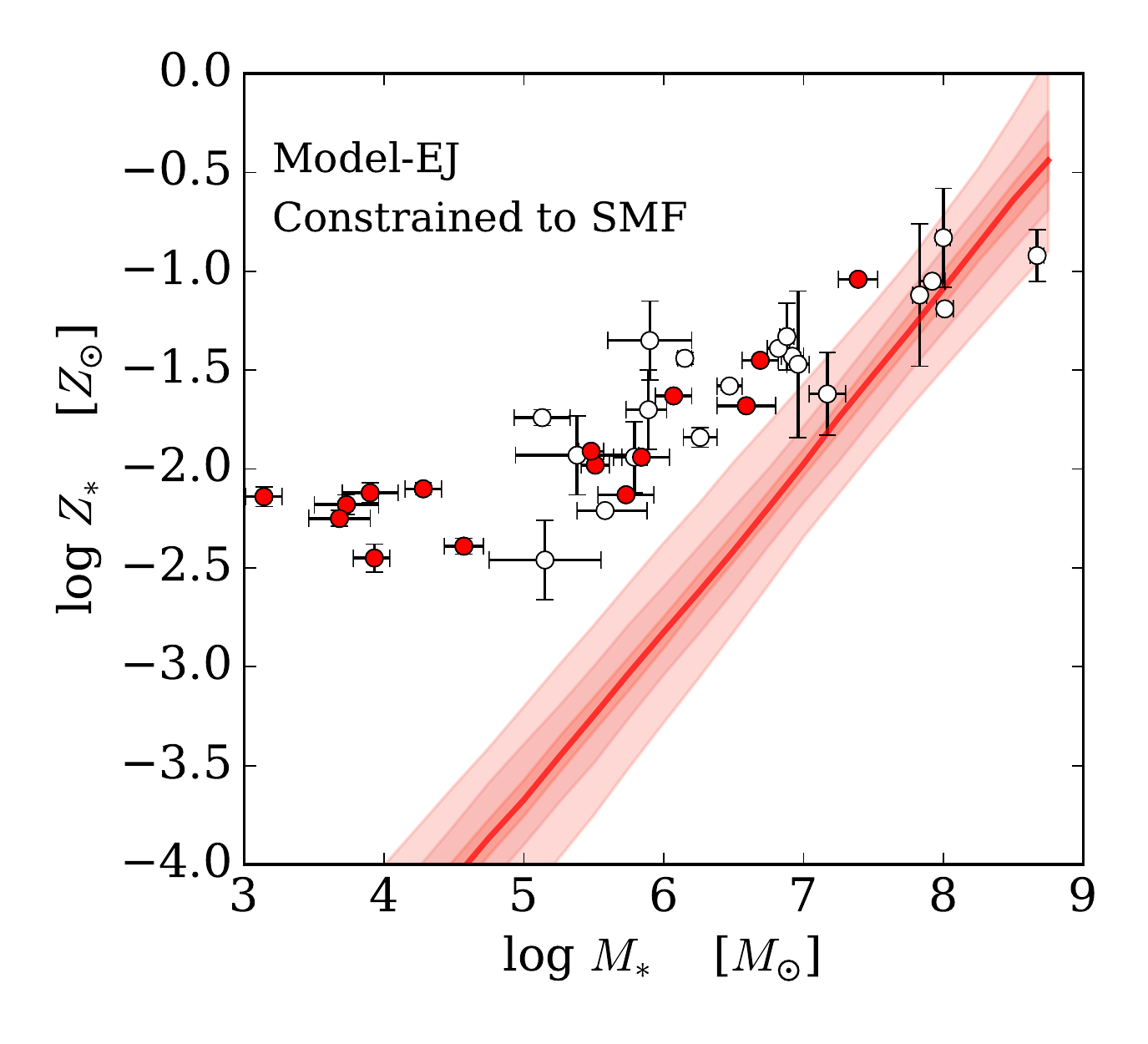}
\includegraphics[width=0.3\textwidth]{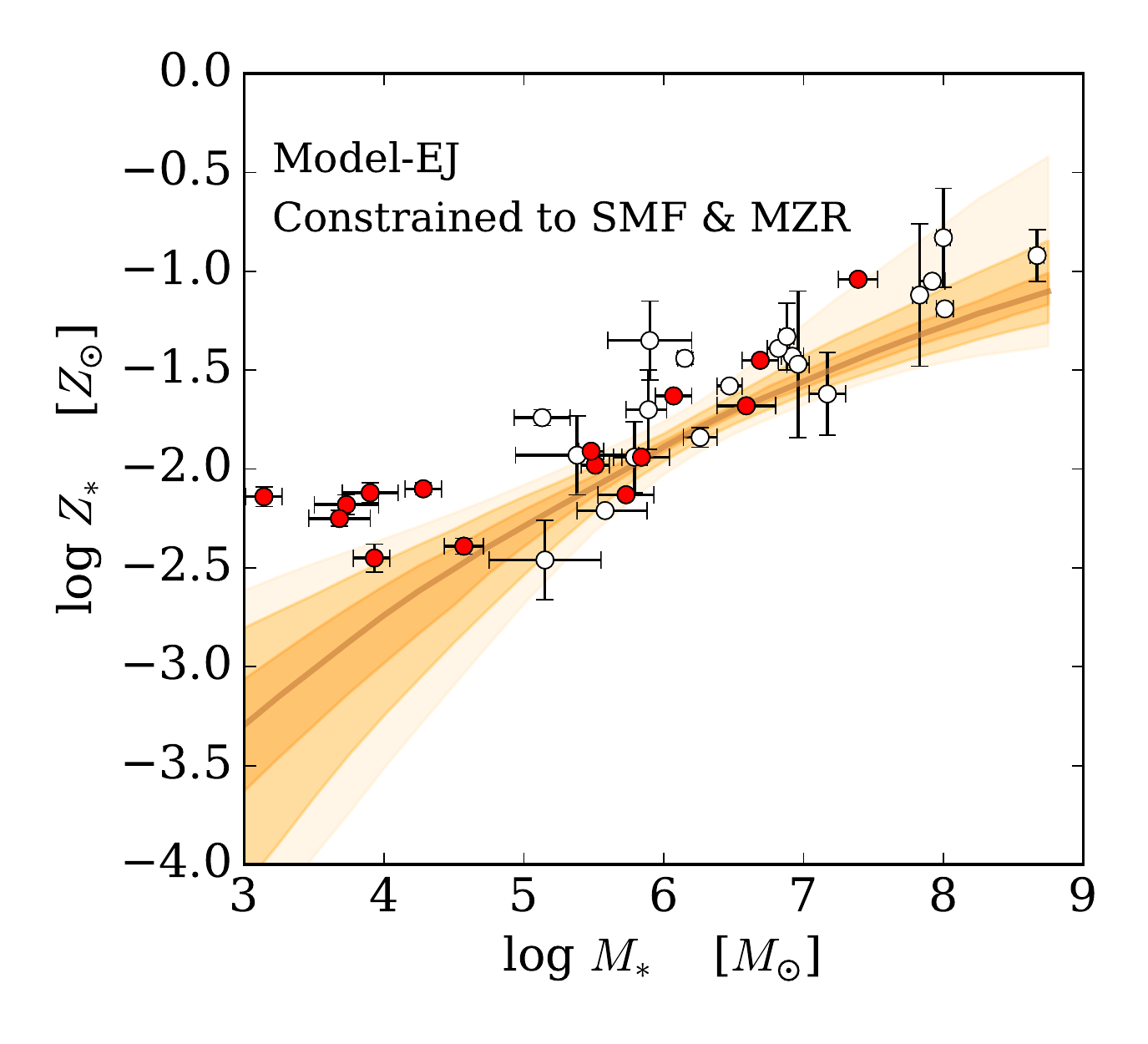}
\includegraphics[width=0.3\textwidth]{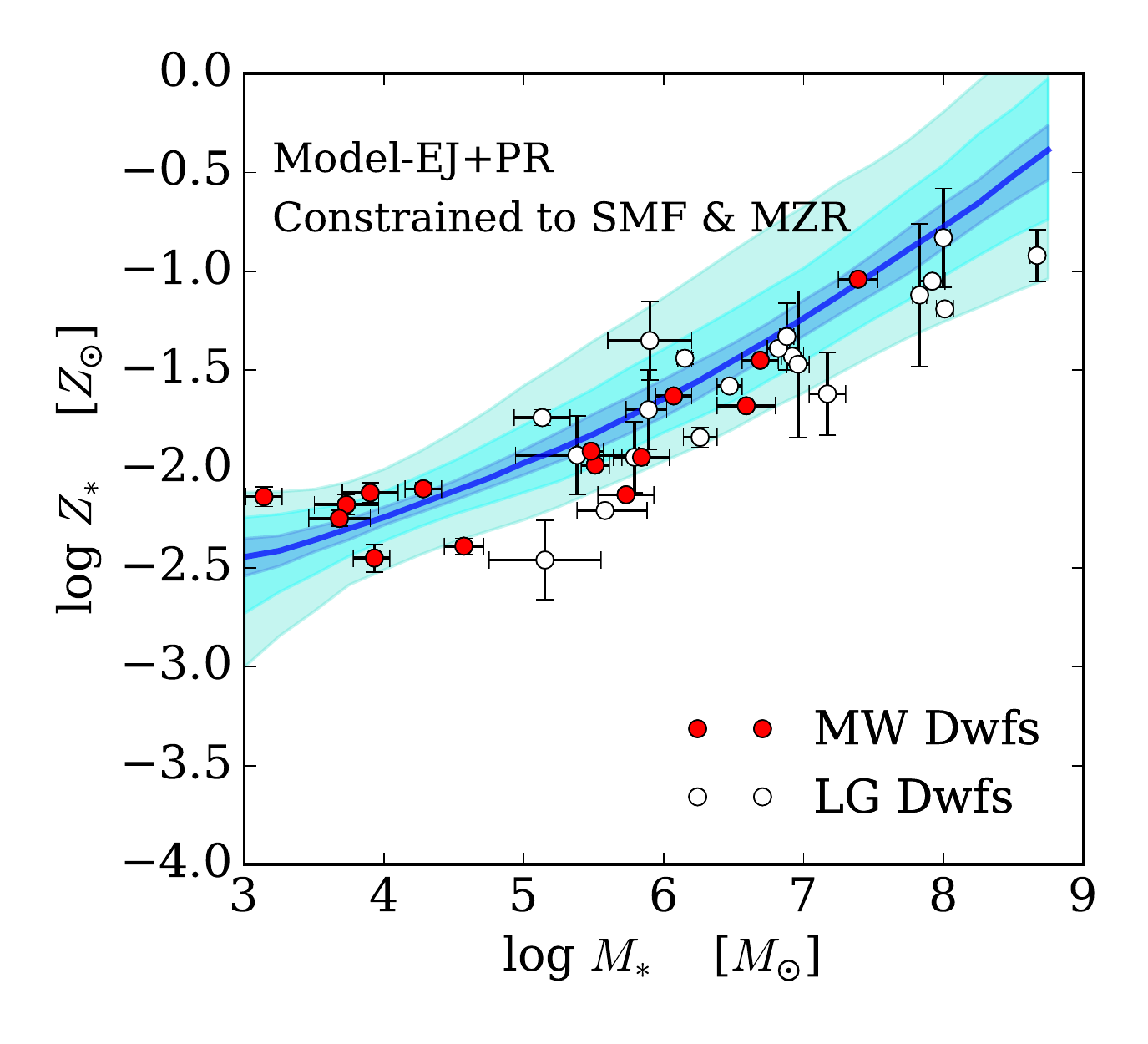}
\caption{Upper: the posterior predictive distribution of the MW satellite stellar mass function predicted by the three models. 
The red circles denote the observational data from \citet{McConnachie2012a}. 
Lower: the posterior predictive distribution of the stellar-phase metallicity as a function of stellar mass of the same models. 
The circles denote observational data from \citet{Kirby2013a}. 
The red filled circles are for the MW dwarfs, and the open circles are for dwarf galaxies in the Local Group but outside the MW. 
Only the MW dwarfs are used as a data constraint in this paper. 
The left column shows the predictions made by the ejective feedback model that is constrained only 
to the MW SMF as presented in \citet{Lu2016a}. 
The middle column shows the predictions made by the ejective model that is constrained to both 
the SMF and the MZR of MW dwarfs. 
Note that the model actually only fits the MZR marginally well, but fails to match the SMF. 
The right column shows the extended model including both ejective and preventive feedback and fit
both the SMF and the MZR of MW dwarfs remarkably well. 
In each panel, the color bands from dark to light encompass the 20\%, 50\%, and 80\% predictive distribution.
}
\label{fig:pred}
\end{center}
\end{figure*}

By marginalizing the posterior distribution of model parameters, we show the predictive distributions for the stellar mass function (upper panels) and the mass--metallicity relation (lower panels) in Figure \ref{fig:pred}. 
In each panel, the bands with different intensity of different color show the 20\%, 50\%, and 80\% predictive distribution.

First, the panels in the left column of Figure \ref{fig:pred} show the predictions of the ejective model constrained to only the MW satellite stellar mass function. 
As discussed in \citet{Lu2016a}, when the model is constrained to the MW satellite stellar mass function, it systematically predicts a steep mass-metallicity relation and significantly underpredicts metallicity for low-mass satellite galaxies. 
For dwarfs with $M_*<10^{5.5}\msun$, the model underpredicts the stellar-phase metallicity by a factor of a few. 
It leaves a question to be answered --- can ejective models simultaneously reproduce the SMF and the MZR of the MW satellite galaxies if we constrain the free parameters using both observational data sets? 

We answer this question by taking both the stellar mass function and the mass--metallicity relation to constrain the model. 
The MCMC exhaustively explores the parameter space and samples the posterior probability distribution of the free parameters under the constraints of both data sets. 
By marginalizing the posterior, we produce the predictive distribution for the MW SMF and MZR, and find that the model is not able to fit both data sets simultaneously.
As we show in the middle column of Figure \ref{fig:pred}, the model predicts too many satellite galaxies with stellar mass below $10^{7}\msun$, resulting in a satellite stellar mass function significantly steeper than observed.
On the other hand, the fit to the mass--metallicity relation is largely improved. 
For galaxies with $M_*>10^5\msun$, the model recovers the shallow slope of the mass--metallicity relation. 
For lower stellar masses, however, the model still tends to predict a steeper slope for the mass--metallicity relation. 
These results suggest that the fit is more strongly influenced by the mass--metallicity relation than the mass function. 
Nevertheless, it is clear that the ejective model family we explore in this paper cannot simultaneously fit both the mass function and the mass--metallicity relation even when a large range of the parameter space is explored.  
Using the result of the flexible model with MCMC, we conclude that the failure of the ejective model is generic but not specific to particular choices of parameter value, and confirm the previous results that additional physics other than outflow rates is needed to resolve the discrepancy in low-mass galaxies \citep[e.g.][]{Font2011a, Henriques2013a, Hou2016a}.

Finally, the right column of Figure \ref{fig:pred} shows the predictions of the extended model that include both ejective and preventive feedback and is constrained to the MW SMF and MZR. 
When preventive feedback is included, the model can fit both data sets remarkably well. 
The constrained model recovers the observed stellar mass--metallicity relation over the entire range of $M_*$ while simultaneously reproducing the stellar mass function. 

We remind the readers that the preventive and ejective forms of feedback are implemented very differently in the model. 
Preventive feedback reduces the accretion of baryons into dark matter halos, and ejective feedback removes a fraction of metal mass from galaxies, resulting in a simultaneous fit to both the SMF and the MZR. 
The success of the phenomenological model suggests that both modes of feedback are required to accurately model the Milky Way's satellites. 
Expulsion of metals via outflows is needed to match the MZR, and suppression of accretion is necessary to reduce the fuel for star formation and reproduce the SMF.

\subsection{Strength of Preventive Feedback}\label{sec:strength}
\begin{figure*}[htbp]
\begin{center}
\includegraphics[width=0.95\textwidth]{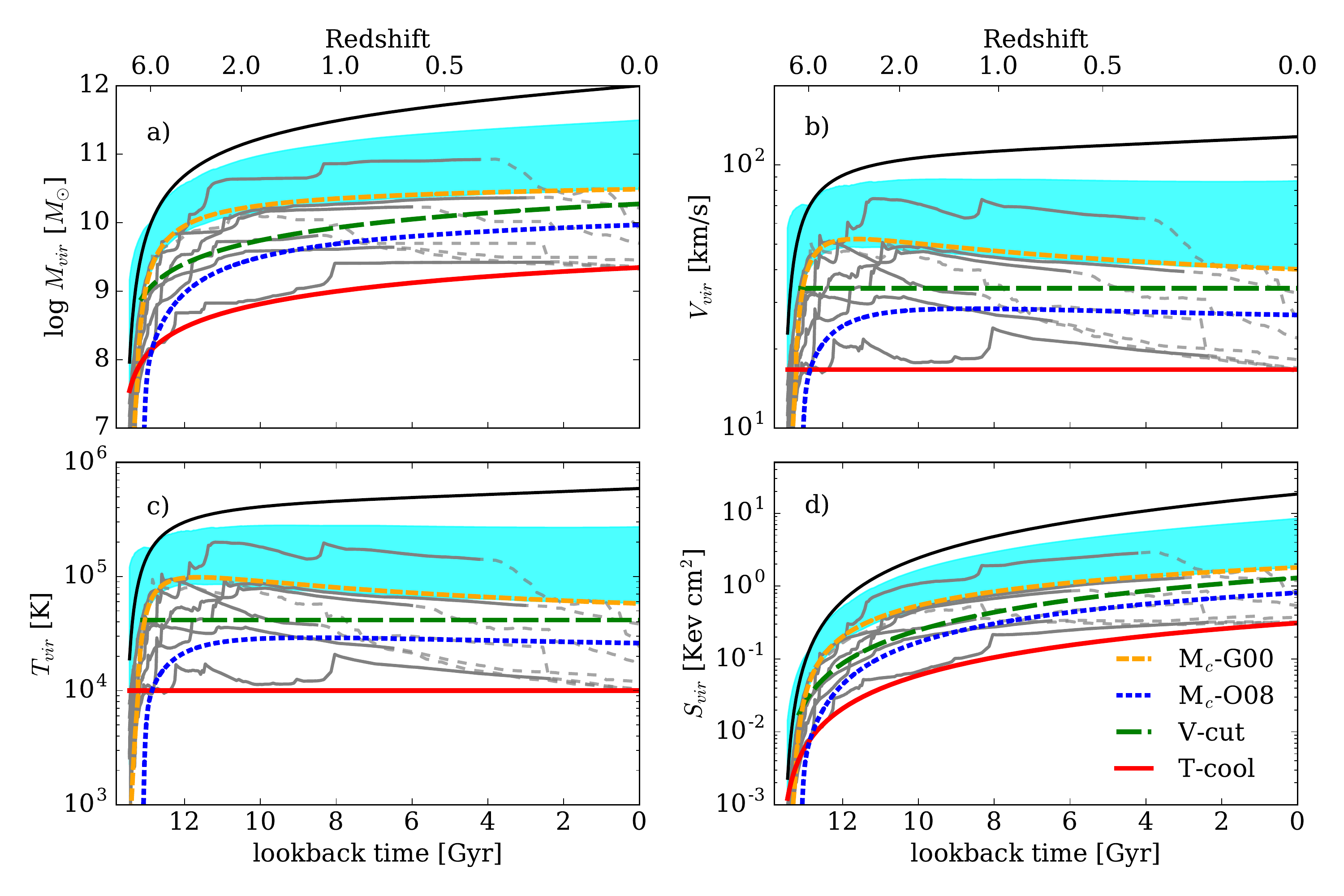}
\caption{
The histories of halo virial mass ($M_\text{vir}$, panel \textbf{a}), virial velocity ($V_\text{vir}$, panel \textbf{b}), virial temperature ($T_\text{vir}$, panel \textbf{c}), and the corresponding virial entropy ($S_\text{vir}$, panel \textbf{d}), as functions of lookback time. 
The gray lines show 7 subhalos randomly selected from all subhalos in our simulations that are more massive than $2\times 10^9\msun$ at $z=0$. 
The solid segment of each line denotes the regime where the halo is a distinct halo in the field, and the faded dashed segment denotes the regime when the halo is identified as a subhalo of another halo. 
The solid black line in each panel shows a smoothed mass assembly history of a typical MW-size host halo. 
The red line shows the atomic cooling limit, corresponding to $T=10^4$K. 
The green dashed line shows the critical accretion limit proposed in \citet{Font2011a}, i.e., $V_{\rm vir} = 34 \kmps$ since $z=10$ to mimic a strong global and local reionization in their model. 
The blue dotted line shows the fitting function provided by \citet{Makiya2016a} for the critical accretion mass affected by reionization as a function of redshift predicted in the hydro simulations of \citet{Okamoto2008a}.
The orange dashed line shows the filtering mass predicted by the reionization model of \citet{Gnedin2000a}. 
The cyan band covers 80\% of the posterior range of the characteristic prevention mass scale in the preventive feedback model, at which mass scale halos can only accrete half of the cosmic baryon fraction.}
\label{fig:subhist}
\end{center}
\end{figure*}

In Figure \ref{fig:subhist}, we demonstrate how the constrained preventive model affects the satellite galaxies. 
We plot the assembly history of 7 subhalos randomly selected from subhalos of the simulated MW host with a final mass $M_{\rm sub}>2\times 10^9\msun$ at $z=0$. 
We use this random selection to eliminate host-to-host variation. 
Among the four panels, panel {\bf a} shows the halo mass assembly histories; 
panel {\bf b} shows the virial velocities; panel {\bf c} shows the virial temperature; and panel {\bf d} shows the virial entropy as a function of time (redshift). 
Each subhalo is denoted by a gray line in the figure, with the solid line denoting the regime when the halo is a distinct halo in the field and the faded dashed line denoting the regime since the halo is first time accreted into another halo. 
The virial velocity is defined as the halo circular velocity at the virial radius. 
The virial temperature of a halo is related to the circular velocity as $T_{\rm vir} = 36.7 \left(V_{\rm c}/\kmps \right)^2 K\,$. 
Following \citet{Lu2015a}, we define the virial entropy as 
\begin{equation}\label{equ:svir}
S_{\rm vir}=\frac{T_{\rm vir}}{n_{\rm vir}^{2/3}}\,,
\end{equation}
with $T_{\rm vir}$ the virial temperature of the halo, and $n_{\rm vir}$ the mean gas particle number density of a virialized halo assuming the cosmic baryon fraction $f_{\rm b,0}$. 
These quantities have meaningful impact on halos only when they are distinct halos (represented by the solid part of the gray lines in Figure \ref{fig:subhist}). 
When the halo becomes a subhalo, its mass and radius take the values assigned by the \textsc{Rockstar} halo identifier \citep{Behroozi2013b}. 
We still use these values to compute the "virial velocity", the "virial temperature" and the "virial entropy", but they no longer have the same physical meaning as for distinct halos. 
We also note that after a halo becomes a subhalo, it is not subject to the preventive feedback but completely loses its halo gas. 
Although it stops accreting new gas, star formation can still continue until the existing cold gas is exhausted. 
The figure shows that the masses of those halos increase rapidly in the first $\sim 2$ Gyrs of the universe (before $z\sim 2$). 
At late times, most of the subhalos lose their mass due to tidal stripping as they are affected by the tidal field of the MW host. 
As one can see, MW subhalos that host classical dwarfs are accreted into another halo typically between $z=1.5$ and 0.5, consistent with the results of \citet{Wetzel2015a}. 

We also show predictions from a handful of models that affect galaxy formation by preventing baryons from cooling or collapsing into halos. 
The red line in each panel shows the atomic cooling limit, corresponding to $T=10^4$K (red line). 
Halos with a virial temperature below the atomic cooling limit are not able to cool baryons to form a galaxy in its potential well unless positive feedback of reionization is considered \citep[for instance, ionizing UV radiation promoting H$_2$ formation, see][]{Ricotti2002b, Bovill2009a}. 
As we can see in the figure, the atomic cooling limit only affects halos when their circular velocities are lower than $17\kmps$.
It is clear that the atomic cooling limit itself is not sufficient as a preventive process to affect most of the observed classical dwarfs.

The other process is the heating due to UV photoionization (photoheating, for short), which not only offsets cooling losses, but also prevents low-mass halos from accreting their full complement of baryons. 
We show the characteristic mass scales in two photoheating models. 
First, we show the characteristic halo mass introduced in the model proposed by \citet{Font2011a}. 
In the model, the authors proposed a model in which halos with $V_{\rm c} \leq 34{\rm km\, s^{-1}}$ cannot cool baryons due to the combination of global reionization and the photoheating in the Local Group for $z<10$.  
We show this scale by a green dashed line in each panel of Figure \ref{fig:subhist}.  
This scale is higher than the atomic cooling scale but has a similar behavior, as it corresponds to a constant virial temperature threshold of $4\times 10^{4}\,{\rm K}$. 
The second photoheating model we consider is the "filtering mass" proposed by \citet{Gnedin2000a} based on linear perturbation theory (orange dashed lines in Figure~\ref{fig:subhist}). 
The model predicts that halos with masses lower than some characteristic mass (the filtering mass) accrete baryons at a significantly reduced rate. 
The filtering mass is higher than the \citet{Font2011a} model, predicting a stronger prevention for halos with higher masses. 
In the \citet{Gnedin2000a} model, the baryon fraction scales with halos as $f_{\rm b} \propto M^{3}$, while the \citet{Font2011a} model imposes a sharp truncation. 
Therefore, the \citet{Font2011a} model has a stronger effect on halos with lower masses. 
Moreover, \citet{Okamoto2008a} derived a model for the suppression of baryonic accretion into halos using hydrodynamical simulations.
As shown in Figure \ref{fig:subhist}, the critical mass scales predicted by the simulations are considerably lower than the model adopted here based on \citet{Gnedin2000a}, suggesting that considering the \citet{Okamoto2008a} model will not affect the preventive model studied in this paper.

Lastly, the cyan band in Figure \ref{fig:subhist} shows the preventive model constrained by the MW satellite galaxy stellar mass function and the mass--metallicity relation in this work. 
The bands show the characteristic halo mass, circular velocity, temperature, and virial entropy at which only half of the baryon mass can collapse into the halo as a function of redshift. 
The band covers 80\% of the posterior distribution of the constrained model.  Below this characteristic halo mass the baryon fraction in the halos scales as $f_{\rm b} \propto M_{\rm vir}^{\beta_{\rm pr}}$, as described in Eq.\,(\ref{equ:fb}).  
In general, the model requires that the characteristic prevention mass scale increases with time. 
At early times ($z>6$), the mass scale can be lower than $10^9\msun$, and becomes higher ($>10^{10}\msun$) at late times.  
The corresponding halo circular velocity is above $40\kmps$, and the virial entropy is  a few $\text{KeV} \, \text{cm}^2$. 
With this level of prevention, most of the MW subhalos, except a couple of the most massive ones, are generally affected by preventive feedback.  
Interestingly, this level of entropy is consistent with the level needed to match the star formation histories of field galaxies with mass equal to or lower than the MW \citep{Lu2015a}. 
In addition to the normalization, the data require preventive feedback that varies with redshift in a similar way as the mass assembly history of the MW host halos (shown as black solid line in Figure \ref{fig:subhist}), suggesting that the evolution of the strength of preventive feedback may be related to the formation of the host halo or its central galaxy.

\section{Discussion and Conclusions}\label{sec:conclusions}
In this paper, we have explored the extensive parameter space of a flexible semi-analytic galaxy formation model within the cosmological evolution of $\Lambda$CDM dark matter halos. We have demonstrated that the ejective feedback model, where feedback is captured only in the form of strong outflows, fails to recover the stellar mass function and the mass--metallicity relation of MW dwarf galaxies simultaneously. 
The strong outflows required to suppress star formation in low-mass halos expel too much mass in the form of metals from low-mass galaxies, resulting a steep mass--metallicity relation that underpredicts the stellar-phase metallicity of MW dwarf galaxies. 

This result is in agreement with the hydrodynamic simulations of \citet{Torrey2014a}, who pointed out that low-mass galaxies need to retain large fraction of metals produced in star formation to avoid predicting a mass--metallicity relation that was too steep relative to observations. 
Recent high-resolution hydrodynamic simulations can also achieve remarkable agreement with the observed mass--metallicity relation over a large stellar mass range when a considerable fraction of metals are retained in the inner halos \citep{Christensen2016a, Ma2016a, Muratov2017a, Angles-Alcazar2017a}.
On the other hand, our inferences show that adopting a smaller mass-loading factor for low-mass halos, so that the galaxies can retain enough metal mass to match the mass--metallicity relation, results in baryonic mass fractions in low-mass subhalos that are too high, thereby predicting a significantly steeper stellar mass function than is observed.
In summary, our inference shows that the ejective feedback model with metal-enriched outflow fails to match the two primary data sets even when many relevant parameters are allowed to vary in a large parameter space. 
If outflow is the primary feedback that suppress baryon mass in low-mass halos, the net outflow has to be metal-deficient.

The difficulty in fitting the data forced us to implement a preventive feedback model into the SAM. 
In this model, a fraction of baryons associated with the dark matter that constitutes the gravitational potential of the dwarf galaxies is prevented from collapsing into the low-mass halos. 
The prevention is increasingly important for lower-mass halos, resulting in a decreasing baryonic-to-dark matter mass ratio for lower halo masses to match the stellar mass function of dwarf galaxies.  
At the same time, moderate ejective feedback is still needed to expel a fraction of metal mass that is mixed into the ISM from galaxies to match the mass--metallicity relation. 
The model with the combination of both preventive and ejective feedback can simultaneously match the stellar mass function and the mass--metallicity relation of MW dwarf galaxies. 
The success of the combined model, coupled with the failure of each model individually, suggests that two different feedback mechanisms are both needed to explain the two pieces of data. 
Specifically, a moderate outflow is needed to match the mass--metallicity relation, and a strong preventive feedback is responsible for governing the low star formation efficiency in low-mass halos. 

This result is in agreement with a number of previous studies. 
Using different SAMs, \citet{Font2011a} and \citet{Hou2014a, Hou2016a} found that lowering the mass-loading factor results in an increased metallicity \citep[also see][]{Guo2016a} and enhanced reionization, which works in the same way as the preventive model we explored in this paper, is needed to suppress star mass in the MW subhalos. 
Similar conclusion about the effect of lowering the mass-loading factor on galaxy metallicity is also reached in hydrodynamic simulation studies  \citep[e.g.,][]{Crain2015a}. 
More recently, using high-resolution hydrodynamical simulations, \citet{Christensen2016a} showed that preventive feedback significantly suppressed baryon accretion in low-mass halos. 
We also note that strong prevention leaves room for metal-enriched outflow (e.g., \citealt{Dalcanton2007a}) to be consistent with the observational constraints. 

The comparison between the constrained preventive model and existing reionization models suggests that a higher degree of prevention than normal reionization, similar to the \citet{Gnedin2000a} filtering mass implementation, is needed to fit the data. 
\citet{Benson2002b} also found that global reionization implemented using the \citet{Gnedin2000a} formalism has a relatively mild effect.
Similarly, using parameter sensitivity analysis on a SAM, \citet{Gomez2014a} found that a stronger prevention than typical reionization models can provide is needed to match the metallicity function of MW satellite galaxies. 
In the study, the authors found that only $\sim 0.05$ of the cosmic baryon fraction of baryons should be allowed to accrete into halos with circular velocity between $30 \kmps$ and $50 \kmps$ in order to match observations.  
If the required prevention suggested in the present paper is due to photoheating of local reionization, the stronger prevention in the Local Group than the global reionization would indicate the importance of considering inhomogeneous reionization \citep[e.g.,][]{Busha2010a, Lunnan2012a, Li2014b}. 

% differential metallicity wind model
We note that there are still large uncertainties in the ejective model and variations in its prescriptions might provide a better match to the data. 
One such variation is to assume that metals are differentially expelled from galaxies or preferentially reincorporated into galaxies with respect to hydrogen (i.e., metal-deficient outflows or metal-enriched inflows). 
For example, if the outflow had lower metallicity than the ISM, it would be possible for the ejective model to predict higher metallicities for low-mass galaxies to match the mass--metallicity relation. 
\citet{Muratov2017a} recently showed that galaxies in hydrodynamic simulations retain a large fraction of metals produced in star formation within their host halos. 
They found that metal-poor gas was ejected from the ISM due to entrainment in outflows generated by the low-intensity star formation.
Alternatively, one can assume that the ejected metals are preferentially reincorporated back into galaxies from the circum-galactic medium (CGM). 
\citet{Li2010a} essentially adopt this type of idea in their SAM by adopting a route to recycle most of the metals produced into newly formed stars through the hot phase. 
In their model, 95\% of newly produced metals are deposited directly into the hot gas for galaxies with a dark matter halo virial mass less than $5\times10^{10}\msun$. 
The metals then eventually fall back to the galaxy and fuel the next episode of star formation, resulting in a shallow metallicity--luminosity relation.

Furthermore, the ejective feedback has other uncertainties that can affect predictions for low-mass galaxies. 
\citet{Henriques2013a,Henriques2015a} showed that when the outflowed mass has a long delay before being reincorporated into low-mass halos, the model can match the evolution of the field galaxy luminosity function remarkably well. 
\citet{Hou2016a} demonstrated a model where the mass-loading factor decreases with time, as suggested by recently hydro-dynamic simulation \citep{Muratov2015a}.
Together with saturated outflows at the low-mass end and a strong local reionization, which are effectively similar to our extended model, the model (named ``EvoFb-LR'' in \citet{Hou2016a}) can achieve a reasonable fit to both the stellar mass function and metallicity-mass relation of MW dwarf galaxies. 
While these prescriptions for outflow and reincorporation have been tested by recent hydro-dynamical simulations, the actual behavior of outflow is still a matter of active debate \citep[e.g.][]{Christensen2016a,Angles-Alcazar2017a}.

We note the important caveat that tidal effects from the MW-mass stellar disk can reduce the number of surviving subhalos \citep[e.g.,][]{Zolotov2012a, Zhu2016a, Wetzel2016a}, which can affect how the ejective model fits the stellar mass function. 
We stress, however, that the mass--metallicity relation is independent from the satellite abundance. Therefore, the requirement for weaker outflow implied by this relation is robust, even if a fraction of subhalos are destroyed by the MW disk. 
Still, it is clear that there is some degeneracy between the preventive feedback and the destruction of subhalos. 
To break this degeneracy, detailed modeling of these processes and quantitative inferences from dwarf galaxies both within and outside the Milky Way will be essential.

Although the model presented here with the addition of preventive feedback can match the data, the physical mechanisms that produce the prevention remain unclear. 
We have taken an extreme model where baryons are prevented from collapsing into low-mass halos, but the assumption has not been fully tested against observation. 
At the current stage, other alternative hypotheses, where baryons still make into the galaxy but are prevented from forming stars by large turbulent pressure or photoheating \citep[e.g.,][]{Forbes2016a}, can produce similar effects in galaxy formation models. 
The data constraints adopted in this paper are not able to distinguish between these scenarios even if models for the relevant physics were implemented in the model. 
Data on the gas and metal content of the ISM and CGM can more directly test if baryons are accreted into the halo or the galaxy. 
Future studies using field low-mass galaxies with cold gas measurements will be needed to further test these models.

In summary, a model including reionization and the ejective feedback of gas with well-mixed metals cannot reproduce both the SMF and MZR.  
However, as we discuss, the parameter space for galaxy formation models is currently poorly constrained, pointing to significant opportunities for future advances. 
Although the metallicity of low-mass galaxies provides a useful constraint for the model, further tests for different feedback processes and metal recycling processes using broader data constraints are needed. 
While models in this work are constrained to local data only, we find that preventive feedback needs to have (largely uncertain) redshift dependence. 
This can be possibly better constrained using the star formation histories of the MW dwarfs \citep[e.g.,][]{Weisz2014a}.
We defer this investigation to a future study, which will help us better understand the origin of the prevention. 
Moreover, the evolution and the gas content of field galaxies with higher masses ($M_*>10^9\msun$) could potentially help break some of the parameter degeneracies by providing independent constraints on the buildup of stellar mass from high redshift to the present day. 
Comprehensive model inferences from compilations of data of field galaxies and dwarf galaxies in the Local Group will improve our understanding of galaxy formation.

%%%%%%%%%%%%%%%%%%%%%%%%%%%%%%%%%%%%%%%%%%%%%%%%%%%%%%%%%%%%%%%%%%%
\acknowledgments
We acknowledge the Ahmanson Foundation for providing computational resources used in this work. 
The zoom-in simulations were performed using computational resources of SLAC National Accelerator Laboratory and of the National Energy Research Scientific Computing Center. We thank the SLAC computational team for their consistent support.   
Support for programs HST-AR-13896, HST-AR-13888, HST-AR-13270, HST-AR-12836, and HST-GO-14734 was provided by NASA through a grant from the Space Telescope Science Institute (STScI), which is operated by the Association of Universities for Research in Astronomy, Inc., under NASA contract NAS 5-26555.
We are grateful to Takashi Okamoto for providing tabulated data for the critical mass as a function of redshift presented in \citet{Okamoto2008a}.
YL thanks Romeel Dav\'{e}, Phil Hopkins, Evan Kirby, Houjun Mo, Josh Simon, and Qingjuan Yu for helpful discussion. 
AW was supported by a Caltech-Carnegie Fellowship, in part through the Moore Center for Theoretical Cosmology and Physics at Caltech.
YYM is supported by the Samuel P.\ Langley PITT PACC Postdoctoral Fellowship, 
and was supported by the Weiland Family Stanford Graduate Fellowship.
ST was supported by the Alvin E.\ Nashman Fellowship.
MBK acknowledges support from NSF grant AST-1517226 in addition to the HST grants listed above. 
%%%%%%%%%%%%%%%%%%%%%%%%%%%%%%%%%%%%%%%%%%%%%%%%%%%%%%%%%%%%%%%%%%%
\bibliographystyle{yahapj}
\bibliography{references}

%%%%%%%%%%%%%%%%%%%%%%%%%%%%%%%%%%%%%%%%%%%%%%%%%%%%%%%%%%%%%%%%%%%

\end{document}